\documentclass[twocolumn,preprintnumbers,superscriptaddress,amsmath,amssymb,longbibliography]{revtex4-2}

\usepackage{graphicx}
\usepackage{dcolumn}
\usepackage{float} 
\usepackage{bbm}

\usepackage{setspace}

\begin{document}

\title{Photon-resolved Floquet theory I:  Full-Counting statistics of the driving field  in Floquet systems }

\author{Georg Engelhardt}
\email{engelhardt@sustech.edu.cn}

\affiliation{Shenzhen Institute for Quantum Science and Engineering, Southern University of Science and Technology, Shenzhen 518055, China}
\affiliation{International Quantum Academy, Shenzhen 518048, China}
\affiliation{Guangdong Provincial Key Laboratory of Quantum Science and Engineering, Southern University of Science and Technology, Shenzhen, 518055, China}

\author{JunYan Luo}

\affiliation{Department of Physics, Zhejiang University of Science and Technology, Hangzhou 310023, China}

\author{Victor M. Bastidas}

\affiliation{Physics and Informatics Laboratory, NTT Research, Inc.,
	940 Stewart Dr., Sunnyvale, California, 94085, USA}

\affiliation{Department of Chemistry, Massachusetts Institute of Technology, Cambridge, Massachusetts 02139, USA}

\author{Gloria Platero}

\affiliation{Instituto de Ciencia de Materiales de Madrid ICMM-CSIC, 28049 Madrid, Spain}

\date{\today}

\pacs{
  }

\begin{abstract}
 	Floquet theory and other established semiclassical approaches are widely used methods to predict the state of  externally-driven quantum systems, yet, they do not  allow to predict the state of the photonic driving field. To overcome this shortcoming, the photon-resolved Floquet theory (PRFT) has been developed recently [Phys. Rev. Research 6, 013116], which deploys concepts  from full-counting statistics to predict the statistics of the photon flux between several coherent driving modes. In this paper, we study in detail the scaling properties of the PRFT in the semiclassical regime. We find that there is an ambiguity in the definition of the moment-generating function, such that different versions of the moment-generating function  produce  the same photonic probability distribution in the semiclassical limit, and generate the same leading-order terms of the moments and cumulants. Using this ambiguity, we establish a simple expression for the Kraus operators, which describe the decoherence dynamics of the driven quantum system appearing as a consequence of the light-matter interaction. The PRFT will pave the way for improved quantum sensing methods, e.g., for spectroscopic quantum sensing protocols, reflectometry in semiconductor nanostructures and  other applications, where the detailed knowledge of the photonic probability distribution is necessary. 
\end{abstract}

\maketitle

\allowdisplaybreaks


\section{Introduction}

Floquet theory and other  semiclassical approaches are very successful in describing quantum matter which is subject to external driving fields. For example, Floquet theory gives deep insight into out-of-equilibrium  quantum phase transitions~\cite{Bastidas2012,Engelhardt2013}, topological phase transitions~\cite{rudner2013anomalous,Benito2014,Engelhardt2016,roy2017floquet,maczewsky2017observation,Schnell2023},  discrete quantum-time crystals~\cite{khemani2016phase,yao2017discrete,else2020longLived,choudhury2021route,Chen2024}, and other effects in which quantum matter is periodically driven~\cite{schweizer2019floquet,cadez2019,NavarreteBenlloch2021,Yan2023,Kolisnyk2024,Kohler2024,Zhang2023}.

\begin{figure*}
	\includegraphics[width=\linewidth]{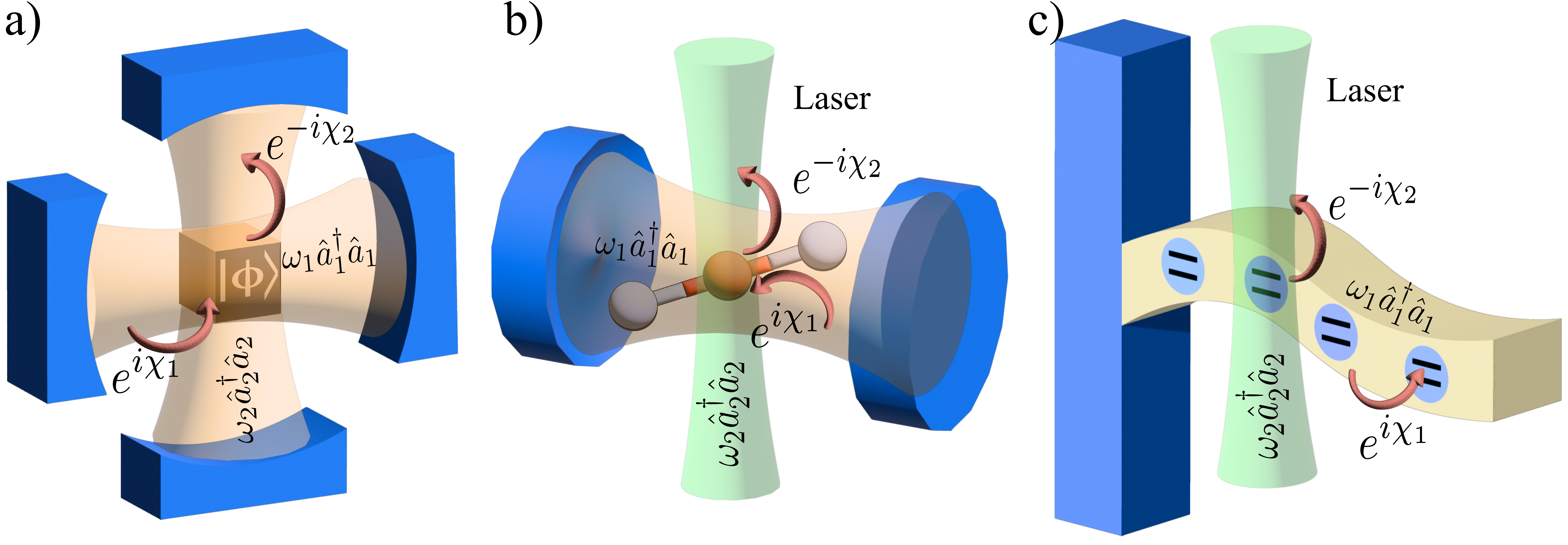}
	\caption{ Possible applications for the photon-resolved Floquet theory (PRFT). a)  illustrates the application to cavity quantum electrodynamics, where a matter quantum system, e.g., an ensemble of cold atoms, interacts with two cavity modes. b) shows the application to polariton chemistry, in which molecules are loaded into an optical cavity and  externally  driven by a laser field. c) depicts a nanomechanical systems where a laser is used to drive the mechanical motion and the motion itself acts as a modulation of the atomic level splitting. In the PRFT, the counting fields $\chi_k$ with $k=1,2$ track the numbers of photons which the cavity (i.e., driving) modes  exchange with the driven quantum system. }
	\label{fig:applications}
\end{figure*}

While this semiclassical approach is a flexible and accurate tool to predict the state of the driven quantum system, Floquet theory per se is not concerned with the state of the driving field. Besides semiclassical methods,  established methods of quantum optics describing the light-matter interaction, such as the celebrated phase-space methods~\cite{Scully1997,Mandel2008}, or input-output theory~\cite{Gardiner2004} can become intractable when dealing with more complex systems.  Other approaches use alternative phase-space methods to gain deeper insights in the dynamics beyond the semiclassical approximation~\cite{Guerin1997}. Recent work investigates the dynamics in Sambe space, essentially a re-quantization of the semiclassical system~\cite{Sambe1973}, to determine the energy transport between photonic modes with different driving frequencies~\cite{Long2021,Crowley2019,Crowley2020}. In Ref.~\cite{Seifert2012}, the quantum-classical transition has been discussed  for a gauge-invariant quantization of the photonic fields. 
 
 However, the aforementioned  methods become  intractable when dealing with multiple photonic driving modes. A recent  work has approached this problem by adapting concepts and methods from full-counting statistics (FCS) of electron transport to photonic systems, in which photons are transported between  multiple photonic driving modes~\cite{Engelhardt2024}. This method has been dubbed photon-resolved Floquet theory (PRFT). In electron transport, the methods of FCS allow to extract the information about the number of electrons having tunneled between distinct electron reservoirs  through a mesoscopic or nanoscopic quantum system by  theoretically monitoring the state of that low-dimensional quantum system~\cite{levitov1996electron,Brandes2004,Schoenhammer2007,Brandes2008}. To this end, so-called counting fields are introduced into the equation of motion of the reduced density matrix of the quantum system, which track the electron exchange with the reservoirs. This flexible theoretical method allows to theoretically explain and control  various  quantum effects, such as the Coulomb blockade~\cite{Bagrets2003}, resonance fluorescence in transport~\cite{Sanchez2007,Sanchez2008},  the reduction of noise~\cite{Ding2024,Xu2023}, just to mention but a few. The methods of FCS have been also applied in the description of phonon transport, based on which the efficiency of thermal devices can be studied~\cite{Restrepo2018,Schaller2018,Cao2023,Wang2024}.
 
 In a similiar fashion as the FCS in electronic systems, the PRFT introduces counting fields into the semiclassical equation of motion of the driven quantum system, which track the photon exchange with the external driving fields. This approach is sketched in Fig.~\ref{fig:applications}(a), which depicts a matter quantum system coupled to two cavity modes, which drive the matter system.  
 Within the PRFT, the dimensionality of the system reduces, as only the state of the driven quantum system has to be simulated, and not the infinite dimensional Hilbert space of the photonic driving fields. Despite its simplicity, numerical benchmarking has shown that the PRFT is very capable to predict the photonic probability distribution of the driving modes~\cite{Engelhardt2024}. Even to such an extent  that it allows to accurately describe a macroscopic light-matter entanglement effect, which has been suggested for a quantum communication protocol~\cite{Engelhardt2024}. This light-matter entanglement effect is illustrated for a two-mode Jaynes-Cummings model in Fig.~\ref{fig:CountingOverview}.

In this paper, we make the concepts and derivations of Ref.~\cite{Engelhardt2024}  precise by investigating in detail the scaling properties of  the PRFT   when approaching the semiclassical limit. Working in Sambe space, we derive an exact expression of the moment-generating function valid for arbitrary photonic initial states. Based on this expression, we discuss how the system approaches a well-defined semiclassical limit and identify the optimal form of the moment-generating function. As we will  explain, there is an ambiguity in the  formulation of the PRFT, as different forms of the moment-generating function will converge to the same probability distributions in the semiclassical limit, which is ultimately related to the central limit theorem of statistics. However, more caution must be exercised for third-and higher-order moments and cumulants: While the leading-order terms in time agree for different versions of the moment-generating function, non-leading-order terms will differ in general. This paper is the first  part of a series of two papers investigating in detail the PRFT. In the companion paper~\cite{Engelhardt2024c}, the PRFT will be applied to open quantum systems based on the investigation of the closed quantum system carried out here. 

Besides quantum optics in cavity systems, we  envision our approach to have impact in many fields of physics such as polariton chemistry \cite{YuenZhou2019,Herrera2020,Ribeiro2018},  and nanomechanical systems~\cite{Ohta2024}, which are illustrated  in Figs.~\ref{fig:applications}(b) and \ref{fig:applications}(c), respectively. 
The theoretical development of an accurate and efficient framework to describe the light-matter interaction has not only a fundamental academic value, but will also pave the way to understand and predict  measurements in spectroscopic experiments in optics~\cite{Panahiyan2023,Dorfman2021,Zhang2022,Gao2021}, and dispersive readout protocols in superconducting circuits~\cite{Blais2021,Zhang2019a}. This might later enable the improvement of quantum sensing protocols, e.g., for the detection of electromagnetic fields or dark matter~\cite{Bloch2022,Engelhardt2024a}, or to speed up the quantum state read-out in superconducting circuits and semiconductor nanostructures~\cite{Vigneau2023,Lecocq2020}. Moreover, already established methods such that the time-dependent density functional theory~\cite{Marques2004} with periodic nuclear motion can also profit from the theoretical methods introduced here.

This paper is organized as follows: In Sec.~\ref{sec:system}, we introduce the light-matter system and the concept of the Sambe space, in which we will carry out our investigation. In Sec.~\ref{sec:fullCountingStatistics}, we rigorously derive an exact expression for the moment-generating function and discuss its semiclassical limit. The  scaling properties of the corresponding cumulants will be illustrated for a two-mode Jaynes-Cummings model. In Sec.~\ref{sec:analysis}, we analyze the convergence of  the probability distribution to a well-defined semiclassical limit, identify the relevant terms in a Floquet expansion, and derive corresponding Kraus operators. In Sec.~\ref{sec:conclusions}, we conclude and discuss the results. Details of the derivations and a discussion about different versions of the moment-generating function can be found in the Appendices.

\section{System}

\label{sec:system}

We consider a generic light-matter interacting quantum system in which a matter subsystem  is coupled to a set of $N_{\text{D}}$ photonic modes, which shall drive the dynamics of the matter system. The Hamiltonian describing such a setup is given by
\begin{equation}
\hat H  = \hat H_{\text{M}} +  \sum_{k=1}^{{N_{\text{D}}}} \omega_k \hat a_k^\dagger  \hat a_k  +  \sum_{k=1}^{{N_{\text{D}}}}  g \hat V_k \left(  \hat a_k^\dagger   + \hat a_k \right),	
\label{eq:hamiltonian:quantum}
\end{equation}
where the Hamiltonian of the matter system is denoted by $\hat H_{\text{M}}$, and the creation and annihilation  operators  quantizing the $k$-th   photonic mode with frequency $\omega_k$  are denoted by $\hat a_k^\dagger$ and  $\hat a_k $, respectively. The operators $\hat V_k $ acting on the matter system specify the light-matter interaction, whose strength is parameterized by $ g$.

In this paper, we consider a separable initial state of the form
\begin{eqnarray}
\left|\psi(0) \right> &=& \left| \phi(0) \right> \otimes \prod_{k=1 }^{{N_{\text{D}}}}  \left|  \alpha_k e^{i\varphi_k} \right> \nonumber \\
 &=& \left| \phi(0) \right>  \otimes  \sum_{\boldsymbol n }  a_{\boldsymbol n} \left|\boldsymbol n \right>,
\label{eq:initalState}
\end{eqnarray}
where $ \left| \phi(0) \right>$ denotes the initial state of the matter system, whereas the states $\left|  \alpha_k e^{i\varphi_k} \right>$ denote coherent states of the photonic modes, i.e., $ \hat a_k \left|  \alpha_k e^{i\varphi_k} \right> = \alpha_k e^{i\varphi_k} \left|  \alpha_k e^{i\varphi_k} \right> $ with real-valued $\alpha_k$ and $\varphi_k$.  In the second line, we have expanded the photonic initial state in the Fock basis $ \left|\boldsymbol n \right>  = \bigotimes_k  \left| n_k \right>   $, where $\boldsymbol n =(n_1,\dots n_{N_{\text{D}}} )$ denotes a vector of photon numbers in each photonic mode labeled by $k$.

Crucially, we assume highly occupied photonic states with large $\alpha_k\gg 1$. This regime allows for a semiclassical description, in which the dynamics of the matter system can be described by the time-dependent Hamiltonian
\begin{equation}
\hat {\mathcal H}_{\boldsymbol \varphi}(t) = \hat H_{\text{M}} +  \sum_{k=1}^{{N_{\text{D}}}}2g\alpha_k \hat  V_k     \cos (\omega_k t - \varphi_k).
\label{eq:def:hamiltonian:semiclassical}
\end{equation}
For notational reasons, we have introduced the vector notation $\boldsymbol \varphi = \left(\varphi_1,\dots, \varphi_{N_{\text{D}}} \right)$. If all frequencies $\omega_k$ are commensurate, the Hamiltonian is time-periodic with period $\tau$, such that it is possible to  apply the celebrated Floquet theory~\cite{Shirley1965}.
According to this theoretical framework, the time-evolution operator of the semiclassical system
\begin{equation}
\hat  {\mathcal U} _{\boldsymbol \varphi} (t) \equiv \hat{\mathcal T }e^{- i \int_{0}^{t}\hat{ \mathcal H_{\boldsymbol \varphi}}(t^\prime ) dt^\prime },
\label{eq:semiclassicalTimeEvolutionOperator}
\end{equation}
where $\hat {\mathcal T}$ denotes the time-ordering operator, can be expanded in terms of the so-called quasienergies $E_{\mu, \boldsymbol \varphi}$ and the time-periodic Floquet states  $\left| u_{\mu,\boldsymbol \varphi}(t) \right> = \left| u_{\mu,\boldsymbol \varphi}(t+\tau) \right>  $ in such that
\begin{equation}
\hat {\mathcal U}_{\boldsymbol  \varphi}(t) = \sum_\mu e^{-i E_{\mu, \boldsymbol \varphi} t}\left| u_{\mu,\boldsymbol \varphi}(t) \right>  \left<  u_{\mu,\boldsymbol \varphi}(0) \right| ,
\label{eq:floquetExpansion}
\end{equation}
which  resembles the expansion of the time-evolution operator for time-independent systems.

However, this is a very drastic semiclassical approach, in which we do not have  access to the  information about the state of the photonic modes. An alternative way to investigate   the high photon number regime can be carried out  by replacing the photonic operators
\begin{subequations}
	\label{eq:def:sambeSpace}
\begin{eqnarray}
\hat a_k^\dagger \hat a_k &\rightarrow& \sum_{n_k=-\infty}^{\infty}  n_k \left| n_k\right> \left< n_k\right|,  \\
\hat a_k^\dagger   &\rightarrow&    \sum_{n_k=-\infty}^{\infty} \alpha_k \left| n_k+1\right> \left< n_k\right| , 
\end{eqnarray} 
\end{subequations}
in the quantum Hamiltonian in Eq.~\eqref{eq:hamiltonian:quantum}, i.e., we extend the Fock space quantum number $n_k$ from $-\infty$ to $\infty$ and neglect the  photon-number dependence of the creation and annihilation operators. The extension to negative photon numbers is justified as the photonic initial state in Eq.~\eqref{eq:initalState} peaks at large photon numbers around $\alpha_k^2\gg1$.
For time-periodic Hamiltonians the replacements \eqref{eq:def:sambeSpace} are equivalent to the representation of the semiclassical Hamiltonian in Eq.~\eqref{eq:def:hamiltonian:semiclassical} in the Sambe space.  Due to this close resemblance, we will thus refer to the representation in Eq.~\eqref{eq:def:sambeSpace} as Sambe space in the following. In doing so, the creation and annihilation operators become translational invariant, which we will use in the following derivations.

\section{Full-counting statistics}

\label{sec:fullCountingStatistics}

To get access to  the full information of the  photonic probability distribution, we take advantage of  the methods of FCS. This allows us to acquire the information about the photonic field by considering only the semiclassical Hamiltonian in Eq.~\eqref{eq:def:hamiltonian:semiclassical}.

\subsection{Moment- and cumulant generating functions}

Here, we first recall the basic defintions and concepts  of the  FCS.  The moment-generating function of the photonic probability distribution  in terms of the real-valued counting fields $\chi_k$ is defined by
\begin{eqnarray}
M_{\boldsymbol \chi} (t) \equiv  \left< e^{-i \sum_k  \chi_k \hat a_k^\dagger \hat a_k   }\right>_t  = \sum_{\boldsymbol n } p_{\boldsymbol n }(t)  e^{-i \boldsymbol \chi\cdot \boldsymbol  n},
\label{eq:def_momentGenFunction}
\end{eqnarray}
where the expectation value is a short-hand for $\left< \bullet \right>_t \equiv \left<\psi(t)\right| \bullet\left|\psi(t)  \right>   $, and $ p_{\boldsymbol n }(t) $ denotes the probability to be in Fock state $\left|\boldsymbol n \right> = \bigotimes_{k=1 }^{N_{\text{D}}} \left| n_k\right> $. Moreover,  we have  introduced the  vector of counting fields  $\boldsymbol  \chi = (\chi_1,\dots,\chi_{N_{\text{D}}})$. Clearly, we can retrieve the photonic probabilities from the moment-generating function via 
\begin{eqnarray}
p_{\boldsymbol n}(t)  &=& \frac{1}{(2\pi)^{N_{\text{D}}}}\int_{-\pi}^{\pi }d\boldsymbol \chi\, M_{\boldsymbol \chi}( t)e^{i\boldsymbol \chi \cdot \boldsymbol n }  ,
\label{eq:probilities}
\end{eqnarray}
i.e., by carrying out a Fourier transformation. Thereby, $\int_{-\pi}^{\pi} d\boldsymbol \chi$ denotes a $N_{\text{D}}$ dimensional integration over all entries of the counting-field vector $\boldsymbol \chi$.  The cumulant-generating function is defined as the logarithm of the moment-generating function, 
\begin{equation}
	K_{\boldsymbol \chi} (t)  = \log M_{\boldsymbol \chi} (t).
\end{equation}
The derivatives of the moment- and cumulant-generating functions 
\begin{subequations}
\begin{eqnarray}
m_l^{(k)} &=&  \frac{d^l}{d\,(-i\chi_k)^l} M_{\boldsymbol \chi= \boldsymbol 0}  \label{eq:moments} \\
\kappa_l^{(k)} &=& \frac{d^l}{d\,(-i\chi_k)^l} K_{\boldsymbol \chi= \boldsymbol 0} 
\end{eqnarray}
\end{subequations}
are the so-called moments and cumulants of the probability distribution of the $k$-th photonic mode, respectively. Likewise, mixed moments and cumulants can be defined by carrying out mixed derivatives of distinct counting fields. According to Eq.~\eqref{eq:def_momentGenFunction} and Eq.~\eqref{eq:moments}, the $l$-th moment is  just a sum over the probabilities weighted by $n_k^l$, where $n_k$ is the photon-number of the $k$-th mode. Thus,  $m_l^{(k)}  = \sum_{\boldsymbol n } p_{\boldsymbol n} n_k^l$.  The cumulants have  interesting physical  interpretations. For instance, the first two cumulants 
\begin{eqnarray}
\overline  n_k   &=& \kappa_1^{(k)}   =  \sum_{\boldsymbol n } n_k  p_{\boldsymbol n },   \nonumber \\
\sigma_k^2 (t) &=& \kappa_2^{(k)} =  \sum_{\boldsymbol n } \left(n_k  - \overline  n_k \right)^2  p_{\boldsymbol n }  
\label{eq:meanAndVariancePRFT}
\end{eqnarray}
denote the mean and the variance of the photonic probability distribution.  Likewise, the third and the forth cumulants, referred to as skewness and kurtosis, carry essential information about the overall shape of the probability distribution. The former  quantifies the asymmetry of the probability distribution, while the latter quantifies the weight in the tails of the distribution in comparison to a Gaussian distribution.

\subsection{Exact moment-generating function}

\label{eq:exact_momGenFct}

In this section, we derive  an exact expression of the moment-generating function defined in Eq.~\eqref{eq:def_momentGenFunction} for a generic photonic initial state in Sambe space, i.e., where the time evolution of the light-matter system is determined by the Hamiltonian in Eq.~\eqref{eq:hamiltonian:quantum} with the photonic operators given in  Eq.~\eqref{eq:def:sambeSpace}. Based on the exact expression, we can  then carry out an adequate semiclassical limit in Sec.~\eqref{sec:semiclassicalLimit}. 

First, we transform the system into an interaction picture defined by the unitary operator
\begin{equation}
\hat U^{(\text{free})}(t)  =e^ { -i \sum_k \omega_k  \hat a_k^\dagger \hat a_k   t },
\end{equation}
such that the exact time evolution operator corresponding to the Hamiltonian in Eq.~\eqref{eq:hamiltonian:quantum} can be written as
\begin{equation}
\hat U(t)  =  \hat U^{(\text{free})}(t) \hat U^{(\text{int})}(t).
\label{eq:timeEvolutionOperator_quantum}
\end{equation}
Thereby, the time evolution operator in the interaction picture is given by
\begin{equation}
\hat U^{(\text{int})}(t)  = \hat{\mathcal T} e^{-i \int_{0}^{t} \hat H^{(\text{int})} (t') dt' } ,
\end{equation}
where
\begin{eqnarray}
\hat H^{(\text{int})} (t)  &=& \hat  H_{\text{M}} \nonumber\\
&+&    \sum_{k=1}^{N_{\text{D}}}  g_k  \hat V_k   \left(\alpha_k e^{i\omega_k t}  \sum_{n_k}  \left| n_k+1\right> \left< n_k\right|   +\text{h.c.} \right).	\nonumber\\
\label{eq:hamiltonian:interaction}
\end{eqnarray}
Notably,  while time dependent, the Hamiltonian in the interaction picture is now translationally invariant in the photon number basis, which facilitates the following derivations. Moreover, it is not hard to see that the time-evolution operator in Eq.~\eqref{eq:timeEvolutionOperator_quantum} and the time-evolution operator in the interaction picture are related as 
\begin{eqnarray}
\left[ \hat U(t) \right]_{\boldsymbol  n,\boldsymbol  n^\prime}  &\equiv&  \left< \boldsymbol n \right| \hat U(t) \left|\boldsymbol n^\prime   \right> \nonumber \\
&=&   e^{-i\boldsymbol \omega \cdot \boldsymbol n t } \left< \boldsymbol n \right|\hat  U^{(\text{int} )}(t)\left|\boldsymbol n^\prime  \right> \nonumber  \\
&\equiv&   e^{-i\boldsymbol \omega \cdot \boldsymbol n t } \hat U^{(\text{int})}_{\boldsymbol n - \boldsymbol  n^\prime }(t),
\label{eq:translationalinvarianceRel}
\end{eqnarray}
where $\boldsymbol \omega  = (\omega_1 ,\dots,\omega_{N_{\text{D}}}) $ is a vector of frequencies.   In  the last line we took advantage of the translational invariance of the interaction Hamiltonian in Eq.~\eqref{eq:hamiltonian:interaction}.

Using Eq.~\eqref{eq:translationalinvarianceRel} and the initial state in Eq.~\eqref{eq:initalState} to evaluate the moment-generating function in Eq.~\eqref{eq:def_momentGenFunction}, we find
\begin{widetext}
	\begin{eqnarray}
	M_{\boldsymbol\chi}(t)   &\equiv& \sum_{\boldsymbol n, \boldsymbol  n_1, \boldsymbol  n_2 }  \left< \phi(0) \right|      \left[ \hat U^\dagger(t) \right]_{\boldsymbol n-\boldsymbol n_{1}, \boldsymbol n }  e^{-i \boldsymbol \chi \cdot  \boldsymbol  n    }  \left[ \hat U(t) \right]_{\boldsymbol n,\boldsymbol n-\boldsymbol n_{2}}   \left| \phi(0) \right>  a_{ \boldsymbol n- \boldsymbol n_{1}}^{*}  a_{ \boldsymbol n-\boldsymbol n_{2}} \\
	&=&  \sum_{\boldsymbol n, \boldsymbol n_1, \boldsymbol  n_2 }   \left<  \     \hat  U^{(\text{int} )\dagger } _{\boldsymbol n_1}(t) e^{-i \boldsymbol \chi \cdot  \boldsymbol  n    }  \hat   U_{\boldsymbol  n_2}^{(\text{int})} (t)     \right>_{t=0}   a_{ \boldsymbol n-\boldsymbol n_{1}}^{*}  a_{ \boldsymbol n- \boldsymbol n_{2}} \nonumber  \\
	&=&\frac{1}{(2\pi)^{3{N_{\text{D}}}} }\iiiint d\boldsymbol \varphi_1 d\boldsymbol\varphi_2 d\boldsymbol\varphi_3d\boldsymbol\varphi_4   \sum_{\boldsymbol n,\boldsymbol  n_1,  \boldsymbol n_2 }   \left<      \hat  U^{\boldsymbol \dagger}_{\boldsymbol \varphi_1} (t)   \hat   U_{\boldsymbol \varphi_2} (t)     \right>_{t=0}  
	a_{\boldsymbol \varphi_3}^{*}  a_{\boldsymbol \varphi_4}  e^{i\boldsymbol  n \cdot(\boldsymbol \varphi_4-\boldsymbol \chi-\boldsymbol\varphi_3) + i \boldsymbol n_1\cdot  (\boldsymbol \varphi_3-\boldsymbol \varphi_1)  - i \boldsymbol n_2\cdot (\boldsymbol\varphi_4-\boldsymbol \varphi_2)  } \nonumber    ,
	\label{eq:tra1:momGenFunction}
	\end{eqnarray}
\end{widetext}
where $\left< \bullet \right>_t \equiv \left<\phi(t)\right| \bullet\left|\phi(t)  \right>   $ is the expectation value in the matter state, and the probability amplitude $a_{ \boldsymbol n}$ was defined in Eq.~\eqref{eq:initalState}. 
In the third equality,  we have introduced the Fourier transformations
\begin{eqnarray}
\hat   U_{\boldsymbol n}^{(\text{int})} (t) &=&   \frac{1}{(2\pi)^{N_{\text{D}}}} \int_{-\pi}^{\pi} d\boldsymbol \varphi\;\hat   U_ {\boldsymbol \varphi } (t ) e^{i   \boldsymbol n \cdot   \boldsymbol \varphi } , \nonumber \\
a_{ \boldsymbol  n} &=&  \frac{1}{(2\pi)^{{N_{\text{D}}}/2}}\int_{-\pi}^{\pi} d\boldsymbol \varphi\;  a_{\boldsymbol \varphi} e^{i  \boldsymbol  n \cdot \boldsymbol  \varphi } .
\label{eq:timeEvolutionOp_fourierTrans}
\end{eqnarray}
The summations over the index vectors $\boldsymbol n,\boldsymbol n_1,\boldsymbol n_2$ generate delta functions, which we can use to evaluate three of the integrals over $\boldsymbol \varphi_1,\dots , \boldsymbol \varphi_4$. In doing so,  the moment-generating function becomes
	\begin{eqnarray}
	M_{\boldsymbol \chi}(t) 
	&=&\int_{-\pi}^{\pi} d\boldsymbol \varphi    \left<      \hat  U^{\dagger}_{\boldsymbol \varphi- \frac{\boldsymbol \chi}{2} } (t)   \hat   U_{\boldsymbol \varphi+\frac{\boldsymbol \chi}{2} } (t)     \right>_{0}  
	a_{\boldsymbol \varphi -\frac{\boldsymbol \chi}{2} }^{*}  a_{ \boldsymbol \varphi+\frac{\boldsymbol \chi}{2}  }  . \nonumber    
	\end{eqnarray}
Crucially, as we show in Appendix~\ref{app:quantumClassicalEquivalence}, we can carry out the exact replacement
\begin{equation}
\hat   U_{\boldsymbol \varphi } (t)   = \hat  {\mathcal U} _{\boldsymbol \varphi} (t),
\label{eq:equivalenceRelationQuantumClassical}
\end{equation}
where the operator on the right-hand side is just the time-evolution operator of the semiclassical system defined in Eq.~\eqref{eq:semiclassicalTimeEvolutionOperator}. Thus, in its final form  the moment-generating function becomes
	\begin{eqnarray}
	M_{\boldsymbol \chi}(t) 
	&=&\int_{-\pi}^{\pi} d\boldsymbol \varphi  \left<      \hat  {\mathcal U}^{\dagger}_{\boldsymbol \varphi-\frac{\boldsymbol \chi}{2}  } (t)   \hat   {\mathcal U}_{\boldsymbol \varphi +\frac{\boldsymbol \chi}{2}  } (t)     \right>_{0}  
	a_{\boldsymbol \varphi -\frac{\boldsymbol \chi}{2} }^{*}  a_{ \boldsymbol \varphi+\frac{\boldsymbol \chi}{2} }   .\nonumber \\
	\label{eq:exactMomentGeneratingFunction}
	\end{eqnarray}
We emphasize that this is an exact expression of the  moment-generating function in Sambe space for arbitrary photonic initial conditions. Noteworthy, it can be evaluated barely in terms of the semiclassical time-evolution operator, and does not require a calculation  of the time-evolution operator in Sambe space, which will be (numerically) very expensive, or even impossible in many systems.

\subsection{Semiclassical limit}
\label{sec:semiclassicalLimit}

Based on Eq.~\eqref{eq:exactMomentGeneratingFunction}, we can take a well-defined semiclassical limit. To this end, we assume the following form of the photonic expansion coefficients in Eq.~\eqref{eq:initalState}
\begin{equation}
a_{ \boldsymbol  n } = \frac{1}{(2\pi )^{\frac{N_{\text{D}}}{4} } \sqrt{\text{det} \boldsymbol \Sigma } } e^{-\frac{1}{4}   \left( \boldsymbol n- \overline {\boldsymbol n} \right)   \boldsymbol \Sigma^{-2}  \left( \boldsymbol n- \overline {\boldsymbol n} \right)   }  e^{i \overline {\boldsymbol \varphi}  \cdot \boldsymbol  n},
\label{eq:coefficientsParameterization}
\end{equation}
where $\overline {\boldsymbol n}  = ( \overline n_1 ,\dots, \overline n_{N_{\text{D}}})  $  is a vector of mean-photon numbers of the photonic modes $k$ and  $\overline {\boldsymbol \varphi}  = ( \overline \varphi_1 ,\dots, \overline \varphi_{N_{\text{D}}}) $ are the corresponding mean phases. The matrix $\boldsymbol \Sigma$ describes the covariance relations of the photonic modes. We thus allow for  more generic initial conditions than the simple product state of photonic modes in the first line of Eq.~\eqref{eq:initalState}.  

The semiclassical limit is defined by letting  the matrix elements of $\boldsymbol \Sigma$ diverge in a controlled manner, meaning that we can define $\boldsymbol \Sigma =\sigma \tilde {\boldsymbol \Sigma}$ with  constant matrix $\tilde {\boldsymbol \Sigma}=\text{const.}$ and diverging scalar factor $\sigma\rightarrow \infty$.  Carrying out the Fourier transformation of the expansion coefficients as defined in Eq.~\eqref{eq:coefficientsParameterization}, we find in the semiclassical limit
\begin{eqnarray}
a_{ \boldsymbol  \varphi} 
&\rightarrow & \frac{\sqrt{ \text{det}\sqrt{2} \boldsymbol \Sigma }  }{(2\pi )^{\frac{N_{\text{D}}}{4} }  }    e^{-\left(  \boldsymbol \varphi  -\boldsymbol {\overline \varphi}     \right)    \boldsymbol \Sigma^2   \left(   \boldsymbol \varphi -\boldsymbol {\overline \varphi}\right)  }e^{i\left(  \boldsymbol  \varphi     -  \boldsymbol {\overline \varphi} \right)  \cdot  \boldsymbol {\overline n}}. \label{eq:expansionCoefficientGausFourier}
\end{eqnarray}
Notably, this expression is not $2\pi$ periodic in $\varphi_k$, yet strongly peaked around $\overline {\boldsymbol \varphi}$, such that the error will be rapidly vanish when evaluating the integral expression in Eq.~\eqref{eq:exactMomentGeneratingFunction} for $\boldsymbol \Sigma\rightarrow \infty$. 

Crucially,  we find that the moment-generating function in Eq.~\eqref{eq:exactMomentGeneratingFunction} exactly factorizes 
\begin{equation}
M_{\boldsymbol  \chi}(t) =M_{\text{dy} ,\boldsymbol \chi}(t)  M_{\boldsymbol  \chi}(0). 
\label{eq:momentGenFkt_factorized}
\end{equation}
Thereby, the second term denotes the initial moment-generating function, which for a  Gaussian state reads
\begin{equation}
M_{\boldsymbol  \chi}(0) =  e^{-i \overline{ \boldsymbol  n } \cdot \boldsymbol \chi  -  \frac{1}{2}\boldsymbol \chi \boldsymbol \Sigma^2 \boldsymbol \chi  }.
\end{equation}
The first term in Eq.~\eqref{eq:momentGenFkt_factorized} incorporates the light-matter interaction and is given by
\begin{multline}
M_{\text{dy} ,\boldsymbol \chi}(t)\\
=\int d\boldsymbol \varphi    \left<     \hat  { \mathcal U}^{\dagger}_{\boldsymbol \varphi- \frac{\boldsymbol \chi}{2} } (t)   \hat   { \mathcal U}_{\boldsymbol \varphi+\frac{\boldsymbol \chi}{2} } (t)     \right>_{0}  
\frac{ e^{-\left(  \boldsymbol  \varphi     -  \boldsymbol {\overline \varphi}   \right)    \boldsymbol \Sigma^2   \left(   \boldsymbol  \varphi     -  \boldsymbol {\overline \varphi}  \right)  }}{\mathcal N}     ,   
\label{eq:dynCumulantGenFct}
\end{multline}
which  we denote  as  dynamical moment-generating function. Thereby, $\mathcal N=     (2\pi )^{\frac{N_{\text{D}}}{2} }/\text{det} \sqrt{2} \boldsymbol \Sigma   $ denotes the normalizing factor.
Importantly, the expectation value in the integrand is weighted by a Gaussian function, which is peaked around $\overline {\boldsymbol \varphi}$. In the semiclassical limit $\sigma\rightarrow \infty$, this Gaussian function converges to a delta function, such that the dynamical moment-generating function in the semiclassical limit becomes
\begin{equation}
M_{\text{dy},\boldsymbol \chi}^{(\text{PRFT})}( t)  =    \left<      \hat{\mathcal U}_{ \boldsymbol {\overline \varphi}  -\frac{\boldsymbol \chi}{2} }^{\dagger} (t)  \hat{\mathcal U}_{\boldsymbol {\overline \varphi} +\frac{\boldsymbol \chi}{2}  }(t)  \right>_{t=0}.
\label{eq:dynCumulantGenFct_PRFT}
\end{equation}
In practice, the convergence of Eq.~\eqref{eq:dynCumulantGenFct} to Eq.~\eqref{eq:dynCumulantGenFct_PRFT} is very rapid. As we show in Appendix~\ref{app:errorAnalysis}, the leading order of the $l$-th cumulant scales with $\kappa_l^{(k)} (t)\propto t^l$. Moreover, the cummulants predicted by Eq.~\eqref{eq:dynCumulantGenFct} and Eq.~\eqref{eq:dynCumulantGenFct_PRFT} are related by
\begin{eqnarray}
\kappa_l^{(k)} (t)= \kappa_l^{(k,\text{PRFT})} (t)  + \mathcal O \left( \frac{t^l}{\sigma^2}   \right) .
\label{eq:errorScaling}
\end{eqnarray}
 Consequently, the relative error is of the order $1/\sigma^2$, which for a coherent state with $\sigma^2 \propto \overline n$ scales as $1/\overline n$. As the moments and cumulants are functions of probabilities $p_{\boldsymbol n}$, which are weighted by factors $n_k^l$, the  error of the higher-order moments and cumulants increases more rapidly than the error of the lower-order ones in the long-time limit. 
 
 \begin{figure*}
 	\includegraphics*[width=1.\linewidth]{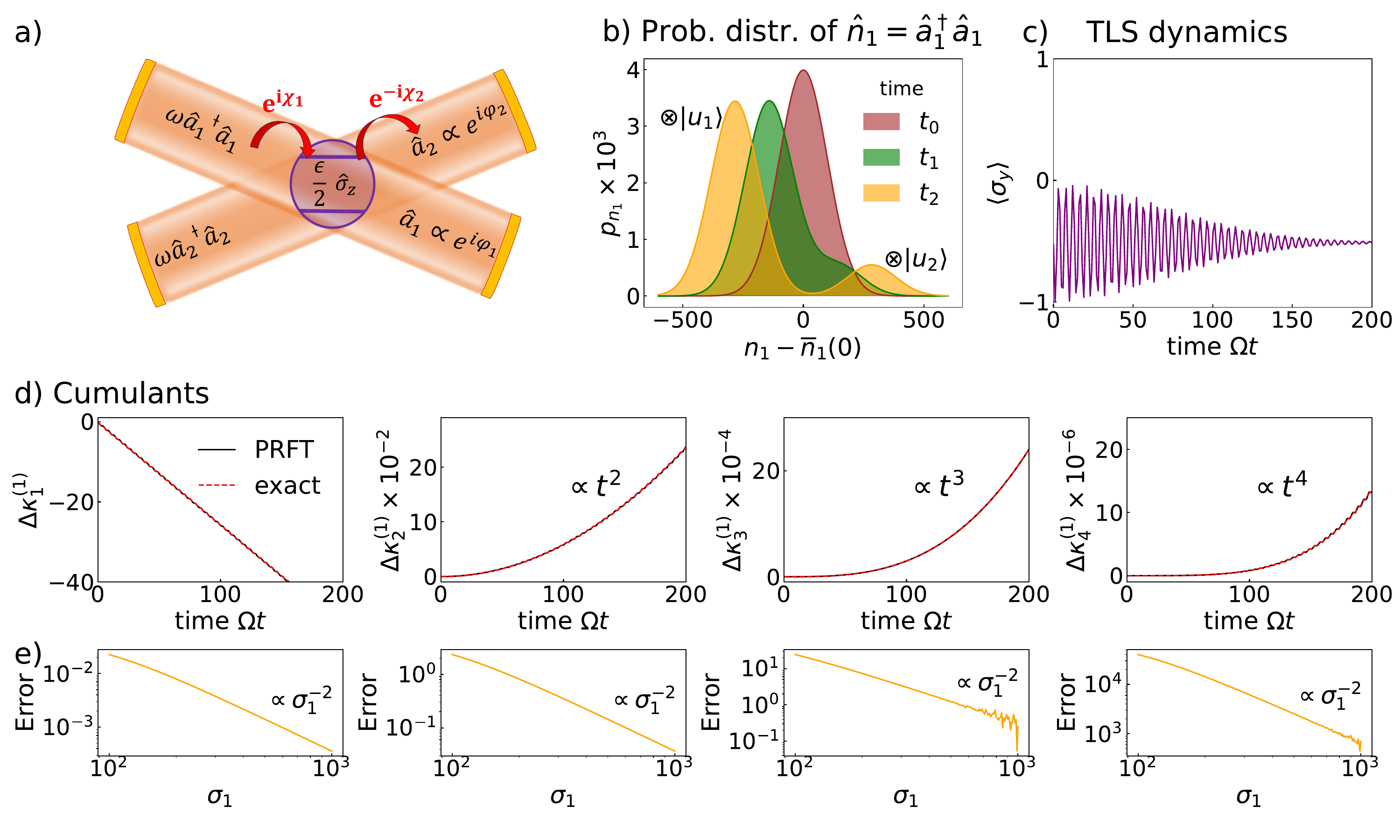}
 	\caption{a) Sketch of the two-mode Jaynes-Cummings model.  b)  Photonic probability distribution of mode $k=1$  for three different times $t_\alpha \Omega =\alpha  \sigma_1$ with $\alpha=0,1,2$, where $\sigma_1^2 = 10^4$ denotes the initial  variance of the photonic probability distribution. The initial state of the two-level system is a superposition of two Floquet states $\left| \phi(0)\right> = c_1\left| u_{1,\boldsymbol \varphi} \right> + c_2\left| u_{2,\boldsymbol \varphi}\right> $ of the semiclassical system, where $c_1 =0.93$ and $c_2 =  0.38$. Parameter are $\Omega_1=\Omega_2 =\Omega$, $\epsilon -\omega = 0.05 \Omega $, and $(\varphi_1,\varphi_2) = (0,\pi/2)$. c) The expectation value  $\left< \hat \sigma_{\text{y}}\right>_{t}$ of the two-level systems shows the decoherence appearing due to  light-matter entanglement. d) First four cumulants of the probability distribution as a function of time calculated using the exact and the PRFT moment-generating functions  in Eq.~\eqref{eq:exactMomentGeneratingFunction} and \eqref{eq:dynCumulantGenFct_PRFT}, respectively. e) Error appearing due to the semiclassical approximation in the PRFT at time $t\Omega=200$.   }
 	\label{fig:CountingOverview}
 \end{figure*}

\subsection{Two-mode Jaynes-Cummings model}

To illustrate the scaling properties of the cumulants, we consider here a two-mode Jaynes-Cummings model governed by the Hamiltonian
\begin{equation}
\hat H_{\text{TMJC} } =   \frac{\epsilon }{2}\hat \sigma_z + \sum_{k=1}^{2} \omega \hat a_k^\dagger\hat a_k  +\sum_{k=1}^{2}     g \left( \hat \sigma_+ \hat a_k  + \hat \sigma_{-} \hat a_k^\dagger    \right)   ,
\label{eq:ham:jcQuantumMM}
\end{equation}
where $\hat \sigma_{\alpha}$ with $\alpha = \left\lbrace \text{x} ,\text{y} ,\text{z} \right\rbrace$ denote the common Pauli matrices. The level splitting is denoted by $\epsilon $ and the light-matter interaction is parameterized by $ g$.  The corresponding semiclassical Hamiltonian  reads accordingly
\begin{eqnarray}
\hat {\mathcal H}_{\boldsymbol \varphi}(t) = \frac{\epsilon }{2}\hat \sigma_z  + \frac{1}{2} \left[\hat  \sigma_+ e^{-i\omega t  }  \Omega(\boldsymbol \varphi )+ \hat \sigma_{-}e^{i\omega t   }  \Omega^*(\boldsymbol \varphi )\right] \nonumber , 
\label{eq:generalizedHamiltonianMMJCmodel}
\end{eqnarray}
where we have defined $ \Omega(\boldsymbol \varphi ) = \sum_{k=1,2} \Omega_k e^{ i \varphi_k   }$ with $\Omega_k =2  g\alpha_k  $ and $\boldsymbol{\varphi}=(\varphi_1,\varphi_2)$. Due to the simplicity of this system, it is possible to find an analytical  expression for the dynamical moment-generating function in the semiclassical limit in Eq.~\eqref{eq:dynCumulantGenFct_PRFT}, which reads
\begin{widetext}
	\begin{eqnarray}
	M_{\text{dy}, \boldsymbol \chi }^{(\text{PRFT})}( t ) &=&  \left[\cos\left( E_{\boldsymbol {\overline \varphi}  - \boldsymbol \chi/2}  t \right) \cos\left( E_{\boldsymbol {\overline \varphi}  + \boldsymbol \chi/2}  t \right)+  \sin\left( E_{\boldsymbol {\overline \varphi}  - \boldsymbol \chi /2}  t \right) \sin \left( E_{\boldsymbol {\overline \varphi}  + \boldsymbol \chi/2}  t  \right) \left<\hat \sigma_{\boldsymbol {\overline \varphi}  -\boldsymbol \chi /2} \hat \sigma_{\boldsymbol {\overline \varphi} + \boldsymbol \chi/2} \right>_0   \right]   \nonumber  \\
	&+&  i \left[\cos\left( E_{\boldsymbol {\overline \varphi} - \boldsymbol \chi/2 }  t \right) \sin \left( E_{\boldsymbol {\overline \varphi}  + \boldsymbol \chi/2}  t \right) \left< \hat \sigma_{\boldsymbol {\overline \varphi}  + \boldsymbol \chi/2} \right>_0 -   \sin \left( E_{\boldsymbol {\overline \varphi}  - \boldsymbol \chi/2}  t \right)  \cos \left( E_{\boldsymbol {\overline \varphi} + \boldsymbol \chi/2}  t \right)\left<\hat \sigma_{\boldsymbol {\overline \varphi}  - \boldsymbol \chi /2} \right>_0  \right], \nonumber  \\
	\label{eq:res:momGenFktJaynesCummingsModel}
	\end{eqnarray}
\end{widetext}
where
\begin{eqnarray}
\hat \sigma_{\boldsymbol \varphi} &=&  \cos\theta_{\boldsymbol \varphi} \hat \sigma_{\text{z}}  + \sin\theta_{\boldsymbol \varphi} \left( \cos\phi_{\boldsymbol \varphi}\hat \sigma_{\text{x} }  + \sin\phi_{\boldsymbol \varphi}\hat  \sigma_{\text{y}   } \right) , \nonumber \\
E_{\boldsymbol \varphi} &=& \frac{1}{2} \sqrt{ \left( \epsilon   -\omega\right)^2   +  \left| \Omega(\boldsymbol \varphi )  \right|^2  }, \nonumber \\%
\tan \theta_{\boldsymbol \varphi} &=& \frac{ \left| \Omega(\boldsymbol \varphi )  \right| }  { \epsilon   -\omega  }, \nonumber \\
\phi_{\boldsymbol \varphi} &=& \text{ arg} \,\Omega(\boldsymbol \varphi ) .
\end{eqnarray}
We note that this expression can be also used to evaluate the integral of the exact expression of the moment-generating function in Eq.~\eqref{eq:exactMomentGeneratingFunction} by replacing $\boldsymbol {\overline \varphi}  \rightarrow  {\boldsymbol  \varphi} $.

In Fig.~\ref{fig:CountingOverview} (d), we compare  the first four cumulants for the photonic statistics of the exact expression (dashed, red) with the ones in the semiclassical limit (solid, black). The initial covariance matrix $\boldsymbol \Sigma = \text{diag} \left(\sigma_1,\sigma_2 \right)$ is diagonal with  entries $\sigma_k = 100$. The initial state of the matter system is  $\left| \phi(0)\right> = 0.93 \left| u_{1,\boldsymbol \varphi } \right>  +  0.38\left| u_{2,\boldsymbol \varphi } \right>    $, where the $\left| u_{\mu,\boldsymbol \varphi } \right>$ denote the Floquet states of the system at time $t=0$. 

We observe that the cumulants indeed diverge  as $\kappa_l^{(1)} (t)\propto t^l$. As explained in Ref.~\cite{Engelhardt2024} and also in more detail in Sec.~\ref{sec:FloquetExpansion}
 this scaling is a consequence of the initial superposition of Floquet states, which generates a macroscopic light-matter entanglement. More precisely, the photonic probability distribution features two well-separated peaks, each one corresponding to one of the two Floquet states, as depicted in Fig.~\ref{fig:CountingOverview}(b). The distance between them increases linearly with time, which dominates the scaling of the higher-order cumulants according to $\kappa_l^{(1)} (t)\propto t^l$.

\begin{figure*}
	\includegraphics*[width=1.\linewidth]{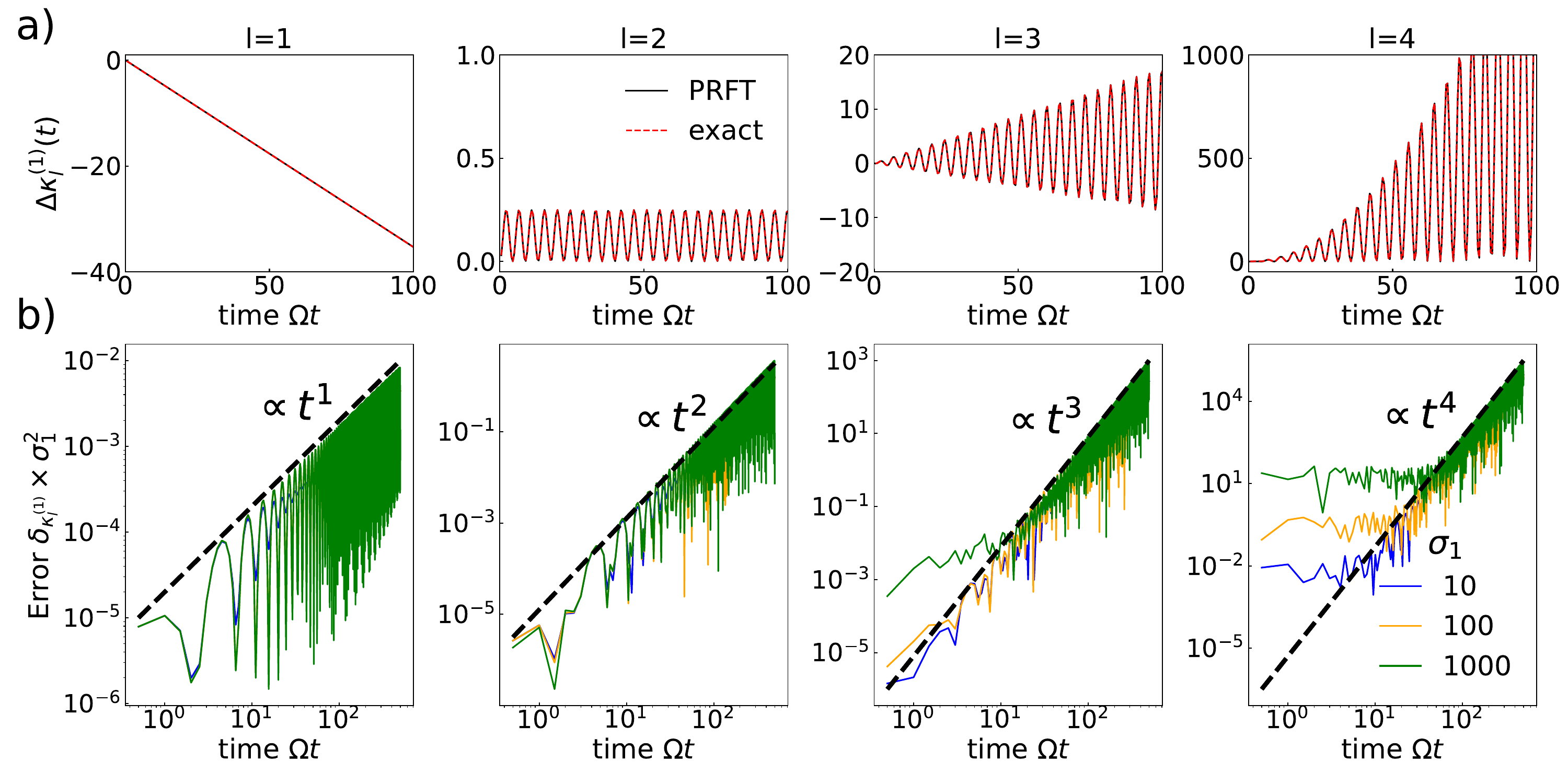}
	\caption{  a) Change $\Delta \kappa_{l}^{(1)}(t) =  \kappa_{l}^{(1)}(t)  -  \kappa_{l}^{(1)}(0) $ of the first four cumulants of the photonic probability distribution of mode $k=1$ in the two-mode Jaynes-Cummings model for an initial Floquet state.  b)  Error $\delta_{ \kappa_{l}^{(1)}} (t) = \left| \kappa_{l}^{(1)}(t)  -  \kappa_{l}^{(1,\text{PRFT})}(t) \right|$   appearing due to the semiclassical approximation in the PRFT for three different initial variances.    Parameters are the same as in Fig.~\ref{fig:CountingOverview}.}
	\label{figErrorAnalysis}
\end{figure*}

In Fig.~\ref{fig:CountingOverview}(e), we depict the error which is caused by the semiclassical approximation. We observe that the error decreases with $\sigma_1^{-2}$ in agreement with the error scaling in Eq.~\eqref{eq:errorScaling}. In Fig.~\ref{figErrorAnalysis}, we investigate the cumulants for an initial Floquet state. As we see in Fig.~\ref{figErrorAnalysis} (a), the prediction of the PRFT  agrees perfectly with the exact result. Now the higher-order cumulants grow only as $\kappa_{l>1}^{(1)} (t)= t^{l-2}$, as the matter system is  initialized in a Floquet state. The error $\delta_{\kappa_{l}^{(1)}} $ between the exact and the PRFT predictions scales as $\delta_{\kappa_{l}^{(1)}} (t)\equiv \left| \kappa_{l}^{(1)}(t)  -\kappa_{l}^{(\text{PRFT}, 1)}(t) \right| \propto  t^{l}/\sigma_1^2$  as shown in Fig.~\ref{figErrorAnalysis}(b).

\section{Scaling analysis}

\label{sec:analysis}

\subsection{Probabilities}

\label{sec:probilities}

According to Eq.~\eqref{eq:errorScaling},  the error of the cumulants due to the semiclassical approximation can rapidly diverge as a function of time. Surprisingly, this divergence is still regular enough such that the exact and approximated probability distributions  converge uniquely when approaching the semiclassical limit for the time scales of interest.  Here, we consider only the probability distribution  of a single photonic mode with mean $\overline n_1 =0$ and variance $\sigma_1^2 =\sigma^2$ to simplify the notation in the following. The generalization to a joint probability distribution is straightforward.

 The light-matter entanglement effect illustrated in Fig.~\ref{fig:CountingOverview}(b) takes place on a time scale of $t\propto \sigma$, i.e., the time  needed that the distance between the two peaks (which grows with $\propto t$) exceeds the width of the peaks (which is of order $\sigma$). To carry out a well-defined semiclassical limit, we thus introduce the scaled time $\tilde t$ via $t =\sigma \tilde t$. The light-matter entanglement effect can now be observed for fixed $\tilde t \propto \mathcal O(1)$. In the same spirit, we also introduce a scaled photon number $\tilde n$ via $n  \equiv \sigma \tilde n$. In doing so, we consider $\tilde n$ as a continuous variable.  The probability distribution of the scaled quantities in the semiclassical limit is then defined via $p^{(\text{sc})}_{\tilde n} (\tilde t) = \lim_{\sigma \rightarrow \infty}\sigma  p_{\sigma \tilde n } (\sigma \tilde t)$. 

The exact cumulants can be distributed as follows
\begin{equation}
\kappa_l(t) =  \kappa_l^{(\text{PRFT})} (t)  +  \frac{ f_l}{\sigma^2} t^{l} ,
\label{eq:cumulantSplitting}
\end{equation} 
where the second term $\frac{ f_l}{\sigma^2} t^{l} $ represents the leading-order terms of the error. Error terms which are of a lower order in time are neglected as they have a minor impact on the probability distribution as compared to the terms scaling with $t^l$.  The cumulants predicted by the PRFT  can be expanded as
\begin{equation}
\kappa_l^{(\text{PRFT})} (t)  =  \delta_{l,2} \sigma^2   +  \sum_{r=1}^{l} 	\kappa_{l,r}^{(\text{PRFT})} (t)  t^r,
\label{eq:cumulantTimeExpansion}
\end{equation}
where  $\sigma^2 $ denotes the initial second cumulant, i.e., the variance.  We recall that we have choosen a vanishing mean photon number $\overline n_1 =0$.  The  terms $\kappa_{l,r}^{(\text{PRFT})} (t) $ are in general time-dependent, but  only of the order of $t^0$ for long times.

In general, the moment-generating function  can be expressed in  terms of the cumulants as 
\begin{eqnarray}
M_\chi(t) &=&   e^{ \sum_l \frac{1}{l!}  (-i\chi)^l  \kappa_l(t) }.
\end{eqnarray}
Using Eqs.~\eqref{eq:cumulantSplitting} and \eqref{eq:cumulantTimeExpansion}, and carrying out the Fourier transformation in Eq.~\eqref{eq:probilities}, we find the  following formal expression for the probabilities
\begin{eqnarray}
p_{n}(t)  &=&  \int_{-\pi}^{\pi} d\chi e^{ i\chi n} 
e^{-\frac{\sigma^2}{2} \chi^2 +  \sum_l \frac{(-i\chi)^l}{l!}  \left( \sum_{r=1}^{l} 	\kappa_{l,r}^{(\text{PRFT})}   t^r   +  \frac{ f_l}{\sigma^2} t^{l}  \right)} \nonumber .\\
\end{eqnarray}
Introducing now the  scaled photon number $n = \tilde n \sigma $ and time $t = \tilde t  \sigma $, and performing the  substitution $y = \chi\sigma$, we can uniquely take the semiclassical limit of the scaled probabilities

\begin{eqnarray}
p_{\tilde n   }^{(\text{sc} )}(\tilde t )  &=& \lim_{\sigma \rightarrow \infty }  \sigma p_{\tilde n \sigma  }(\tilde t\sigma ) \nonumber \\
&=&  \lim_{\sigma \rightarrow \infty } \int_{-\pi}^{\pi}  dy  e^{i \tilde n y  }\nonumber \\
&& \qquad \times e^{-\frac{y^2}{2}  +  \sum_l \frac{(-iy)^l}{l!}  \left( \sum_{r=1}^{l} 	\kappa_{l,r}^{(\text{PRFT})} \tilde   t^r \sigma^{r-l}   +  \frac{ f_l}{\sigma^{2}} \tilde t^{l}  \right) }  	  \nonumber  \\
&=&    \int_{-\pi}^{\pi} dy  
e^{-\frac{y^2}{2}  +  \sum_l \frac{1}{l!}(-iy)^l  	\kappa_{l,l}^{(\text{PRFT})}  \tilde t^l  }  	  e^{i \tilde n y  }  \nonumber  \\
&= & \lim_{\sigma \rightarrow \infty }   \sigma p_{\tilde n \sigma  }^{(\text{PRFT})}(\tilde t\sigma ) .
\label{eq:semiclassicalProbilityDistribution}
\end{eqnarray}
In the second  line, we observe that the error terms $\propto \tilde t^l \tilde y^l / \sigma^2$ vanish in the semiclassical limit. Therefore, we conclude that the photonic probability distribution generated by the exact and PRFT moment-generating functions converge to each other in the semiclassical limit $\sigma\rightarrow \infty$. Likewise, also the non-leading order terms of the PRFT cumulants $\kappa_{l,r}^{(\text{PRFT})}$ with $r<l$ vanish. Thus, the photonic probability distribution in the semiclassical limit is solely determined the leading order terms $\kappa_{l,l}^{(\text{PRFT})}$ of the cumulants predicted via the PRFT.

\begin{figure*}
	\includegraphics*[width=1.\linewidth]{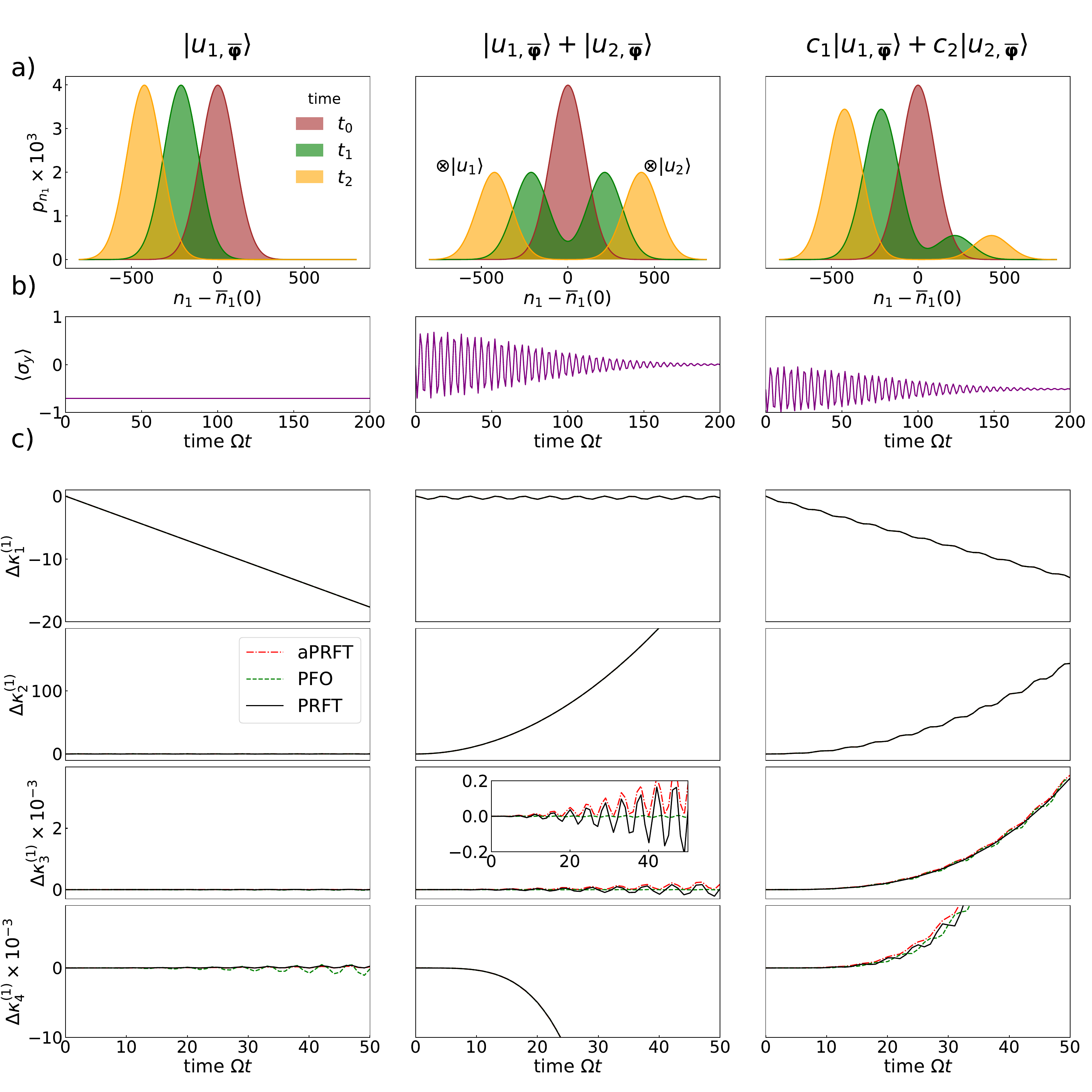}
	\caption{a) Photonic probability distribution in mode $k=1$ of the two-mode Jaynes-Cummings model in the semiclassical regime [Eq.~\eqref{eq:ham:jcQuantumMM}] for  an initial Floquet state (left), a balanced superposition of Floquet states (middle), and an unbalanced superposition of Floquet states with $(c_1,c_2) = (0.93,0.38)$ (right). Parameters are the same as in Fig~\ref{fig:CountingOverview}. b) Expectation value of $\hat \sigma_{\text{y}}$ of the two-level system as a function of time. c) Change of the first four cumulants as  predicted by the PRFT in Eq.~\eqref{eq:dynCumulantGenFct_PRFT} depicted by a black (solid) line. The green (dashed) and red (dash-dotted) lines show the calculation via the photon-flux operator approach [Eq.~\eqref{eq:EOF_momGenFct}] and the alternative PRFT [Eq.~\eqref{eq:alt_momGenFct}], respectively. We observe that the first two cumulants are equal for all three approaches, while for the higher-order cumulants, only the leading order terms agree for all three approaches.   }
	\label{fig:CountingSemiClassial.pdf}
\end{figure*}

\subsection{Floquet expansion}

\label{sec:FloquetExpansion}

In many cases of practical relevance, the frequencies of the driving field are commensurate, such that the semiclassical Hamiltonian in Eq.~\eqref{eq:def:hamiltonian:semiclassical} is periodic in time. In this case, we can express the time-evolution operator in terms of the quasienergies and the Floquet states according to Eq.~\eqref{eq:floquetExpansion}. In doing so, we find
\begin{eqnarray}
M_{\text{dy},\boldsymbol \chi}^{(\text{PRFT})}( t)   &=&  \left<      \hat{\mathcal U}_{ \overline{\boldsymbol\varphi} -\boldsymbol \chi/2 }^{\dagger} (t)  \hat{\mathcal U}_{\overline{\boldsymbol\varphi} +\boldsymbol \chi/2}(t)  \right>_{t=0}\nonumber  \\
&=&  \sum_{\alpha,\beta,\mu,\nu} c_\alpha^* c_\beta  e^{i(E_{\mu, \overline{\boldsymbol\varphi}  -\boldsymbol \chi/2}  -E_{\nu,\overline{\boldsymbol\varphi} + \boldsymbol \chi/2}  )t }\nonumber  \\
&& \times \left<  u_{\alpha,\overline{\boldsymbol\varphi}  }(0)	 \mid u_{\mu,\overline{\boldsymbol\varphi}  -\boldsymbol \chi/2} (0) \right> \nonumber  \\
&& \times  \left<  u_{\mu,\overline{\boldsymbol\varphi}  -\boldsymbol \chi/2} (t)\mid u_{\nu,\overline{\boldsymbol\varphi} + \boldsymbol \chi/2}(t) \right>\nonumber    \\
&& \times \left<  u_{\nu,\overline{\boldsymbol\varphi} +\boldsymbol \chi/2} (0)	  \mid  u_{\beta,\overline{\boldsymbol\varphi} } (0)	  \right>, 
\label{eq:momGenFct_FloquetExpansion}
\end{eqnarray}
where  the $c_\mu$  denote the expansion coefficients of the initial state in the Floquet basis, i.e., $\left|\phi(0)\right> = \sum_{\mu} c_{\mu} \left| u_{\mu,\overline{\boldsymbol\varphi} } (0)  \right> $.
In the previous section, we found that  the probability distribution in the semiclassical limit is solely determined by the leading orders in time of  the cumulants. Using the Floquet expansion in Eq.~\eqref{eq:momGenFct_FloquetExpansion}, it is straightforward to show that 
\begin{eqnarray}
M_{\text{dy},\boldsymbol \chi}^{(\text{sc})}( t)    
&=&  \left< \sum_{\mu }    e^{-i t \boldsymbol \nabla_{\boldsymbol \varphi}  E_{\mu, \overline{ \boldsymbol \varphi} }  \cdot \boldsymbol \chi  }   \left| u_{\mu,\overline{\boldsymbol  \varphi} } \right>  \left<  u_{\mu,\overline{ \boldsymbol  \varphi} } \right| \right>_{t=0}  ,
\label{eq:momGenFkt_semiclassicalLimit}
\end{eqnarray}
where $\left| u_{\mu,\overline{\boldsymbol  \varphi} } \right>=\left| u_{\mu,\overline{\boldsymbol  \varphi} } (0)\right>$, produces the same leading-order terms of the moments and cumulants as the moment-generating function in Eq.~\eqref{eq:momGenFct_FloquetExpansion}.  Thus, the semiclasscial probability distribution is exclusively determined by the first derivatives of the quasieneriges, i.e., $\left( \boldsymbol \nabla_{\boldsymbol \varphi}   E_{\mu ,\overline{\boldsymbol  \varphi}}\right)_k   = \partial_{\varphi_k} E_{\mu,\boldsymbol \varphi = \overline{\boldsymbol  \varphi} } $. 

Using the moment-generating function in the semiclassical limit in Eq.~\eqref{eq:momGenFkt_semiclassicalLimit} to calculate the probabilities according to Eq.~\eqref{eq:probilities}, we find that the time-evolved probability distribution satisfies the relation
\begin{equation}
p_{\boldsymbol  n}(t)  =  \sum_\mu \left| c_\mu  \right|^2 p_{\boldsymbol  n -\boldsymbol { \overline  n}_\mu (t) }(0) 
\label{eq:semiclassicalProbabilityDistribution},
\end{equation}
where we have introduced the vector of the mean photon number change $\Delta \boldsymbol {  n}_\mu (t)   =   t \boldsymbol \nabla_{\boldsymbol \varphi}  E_{\mu, \overline{\boldsymbol  \varphi} }  $. This simple dependence of the probability distribution   on the Floquet states has been already investigated based on an alternative version of the PRFT in Ref.~\cite{Engelhardt2024}, which converges in the semiclassical limit  to the same semiclassical  probability distribution [Eq.~\eqref{eq:momGenFkt_semiclassicalLimit}] as discussed in Appendix~\ref{app:alternativeFCSapproaches}.

 The probability distribution in Eq.~\eqref{eq:semiclassicalProbabilityDistribution} describes an entangled  light-matter state, where the state of the  matter system steers the dynamics of the photonic system.   We illustrate this effect for  the Jaynes-Cummings model in Fig.~\ref{fig:CountingSemiClassial.pdf}(a), which depicts the semiclassical probability distribution as defined in Eq.~\eqref{eq:semiclassicalProbilityDistribution} and evaluated using Eq.~\eqref{eq:semiclassicalProbabilityDistribution}.

 The first column shows the probability distribution for an initial Floquet state of the two-level system. As there is only one coefficient in the Floquet expansion with  $\left| c_{\mu}\right|^2=1$, the initial Gaussian distribution is simply shifted in position towards lower photon numbers $n_1$. Physically, this describes just a flux of photons from mode $k=1$ to mode $k=2$. 
 
 In the second column, the matter initial state is a balanced superposition of Floquet states with $\left| c_{1}\right|^2=\left| c_{2}\right|^2= 0.5$. Each Floquet state $\left|u_{\mu,\overline{\boldsymbol  \varphi}} \right>$ steers the photon transport in a different direction. While for $\mu=1$, the number of photons in mode $k=1$ decreases, for  $\mu=2$ the number increases linearly in time. According to Eq.~\eqref{eq:semiclassicalProbabilityDistribution}, this leads to a superposition of two clearly separated  Gaussian peaks for long enough times, which reflects an entangled light-matter state.
 
  The third column in Fig.~\ref{fig:CountingSemiClassial.pdf}(a) depicts a more generic  initial state $\left| \phi(0)\right>  = c_1 \left| u_{1,\overline{\boldsymbol  \varphi}} \right >  +   c_2 \left| u_{2,\overline{\boldsymbol  \varphi} } \right >   $ with $c_1 =0.93$ and $c_2 =  0.38$, which is the same as in Fig.~\ref{fig:CountingOverview}(b).
  Similar to the balanced superposition before,  the initial Gaussian peak splits into two peaks, each of which is entangled with one of the two Floquet states $\mu$. Accoring to Eq.~\eqref{eq:semiclassicalProbabilityDistribution}, the area under each peak is given by $\left| c_\mu\right|^2$, respectively.

\subsection{ Kraus operators}

In Sec.~\ref{sec:FloquetExpansion}, we have heuristically explained the appearance of the double peak structure in Fig.~\ref{fig:CountingOverview} by the generation of light-matter entanglement. In this section, we investigate this effect more rigorously. 

To this end, we consider first the time-evolved state of the joint light-matter system, which can be formally parameterized as
\begin{eqnarray}
\left| \psi(t) \right>  =  \sum_{\boldsymbol n,\mu } a_{\boldsymbol n,\mu}(t) \left|\boldsymbol n \right>\left| u_{\mu,\overline{\boldsymbol  \varphi}}(t) \right>.
\end{eqnarray}
Thereby, $\left|\boldsymbol n \right>$ denote the Fock states of the photonic modes and $\left| u_{\mu,\overline{\boldsymbol  \varphi}}(t) \right>$ represents the Floquet states of the semiclassical Hamiltonian. Similar as the moment-generating function, the expansion coefficients can be expressed in a semiclassical fashion via
\begin{eqnarray}
a_{\boldsymbol n,\mu}  (t)&=& \sum_{\boldsymbol n_{1}} \left< u_{\mu,\overline{\boldsymbol  \varphi}}(t)  \right|  \left[ \hat U(t) \right]_{\boldsymbol n , \boldsymbol n-\boldsymbol n_{1}}  \left|\phi(0) \right> a_{ \boldsymbol n-\boldsymbol n_{1} } \nonumber   \\
&=&  e^{-i\boldsymbol \omega \cdot \boldsymbol n t }  \sum_{\boldsymbol n_{1} }  \left< u_{\mu ,\overline{\boldsymbol  \varphi}} (t)\right| \hat U^{(\text{int})}_{\boldsymbol n_{1}}(t)  \left|\phi(0) \right> a_{ \boldsymbol n-\boldsymbol n_{1} }  \nonumber  \\
&=&  \frac{ e^{-i\boldsymbol \omega \cdot \boldsymbol n t }}{(2\pi)^{\frac{R}{2} }}  \int_{-\pi}^{\pi} d\boldsymbol \varphi  \left< u_{\mu ,\overline{\boldsymbol  \varphi} }(t)\right| \hat{\mathcal U}_{\boldsymbol \varphi}(t)  \left|\phi(0) \right> a_{\boldsymbol  \varphi  }  e^{i\boldsymbol \varphi \cdot \boldsymbol n }  ,\nonumber \\
\end{eqnarray}
where the defintions of the symbols can be found  in Sec.~\ref{eq:exact_momGenFct}. We emphasize that this  expression is still exact.

Guided by our findings in Sec.~\ref{sec:FloquetExpansion} that the probability distribution is mainly determined by the quasieneriges, we now replace 
\begin{equation}
\hat{\mathcal U}_{\boldsymbol \varphi}(t) \rightarrow \sum_\mu 	e^{-i t \left[E_{\mu, \boldsymbol{ \overline \varphi}}  +  \boldsymbol \nabla_{\boldsymbol \varphi}  E_{\mu, \overline{ \boldsymbol \varphi} }  \cdot  \left(\boldsymbol \varphi - \boldsymbol{ \overline \varphi} \right) \right]  }   \left| u_{\mu,\overline{ \boldsymbol \varphi}}(t) \right>  \left<  u_{\mu, \overline{ \boldsymbol \varphi}  } \right| ,
\label{eq:semiclassicalTimeEvolutionOperatorApprox}
\end{equation}
such that the  expansion coefficients become
\begin{eqnarray}
a_{\boldsymbol n,\mu} (t) &=&  
c_{\mu}  e^{-i( \boldsymbol \omega \cdot \boldsymbol n  +E_{\mu, \overline{ \boldsymbol \varphi}  } ) t } \nonumber\\
 &&\times \int\frac{ d\boldsymbol \varphi }{(2\pi)^{N_{\text{D}}/2}}  e^{-i t \boldsymbol  \nabla_{\boldsymbol  \varphi} E_{\mu, \boldsymbol {\overline \varphi}  }\cdot (\boldsymbol  \varphi-\overline{\boldsymbol \varphi} ) } e^{i\boldsymbol \varphi \cdot \boldsymbol n } a_{ \boldsymbol  \varphi  }  \nonumber  \\
&=&  
c_{\mu}  e^{-i( \boldsymbol \omega \cdot \boldsymbol n  +E_{\mu, \overline{ \boldsymbol \varphi}  } ) t } e^{i   \Delta \boldsymbol n_{\mu}(t) \cdot \boldsymbol {\overline \varphi}    }    a_{ \boldsymbol n -  \Delta \boldsymbol n_{\mu}(t)  }  ,\nonumber \\
\label{eq:expansionCoefficient}
\end{eqnarray}
where $\Delta \boldsymbol n_{\mu}(t) $ has been defined below Eq.~\eqref{eq:semiclassicalProbabilityDistribution}. This expression explicitly shows how the expansion coefficients of the photonic basis are shifted with time depending on the  Floquet state, i.e., how light-matter entanglement is created.

In a similar fashion as the expansion coefficients in Eq.~\eqref{eq:expansionCoefficient}, we can also derive an expression for the state of the matter system conditioned on a particular Fock state:
\begin{eqnarray}
\rho_{\boldsymbol n}(t) &\equiv &  \frac{1}{p_{\boldsymbol n}(t) }	\left< \boldsymbol  n\mid  \psi(t) \right> \left< \psi(t)\mid \boldsymbol  n  \right> \nonumber  \\
&=&  \frac{1}{p_{\boldsymbol n}(t) } \mathcal K_{\boldsymbol n} (t) \rho(0)  \mathcal K_{\boldsymbol n}^\dagger (t),
\end{eqnarray}
with  $\rho(0) =\left|\phi(0) \right>\left<\phi(0)  \right|$ denoting the initial density matrix of the matter system, which has been expressed in terms of the Kraus operators
\begin{eqnarray}
\mathcal K_{\boldsymbol n} (t)   &=&   e^{-i\boldsymbol \omega \cdot \boldsymbol n t }  \int \frac{ d\boldsymbol \varphi}{(2\pi)^R}  \hat{\mathcal U}_{\boldsymbol \varphi}(t)   a_{ \boldsymbol \varphi  }  e^{-i\boldsymbol \varphi \cdot \boldsymbol n },
\label{eq:kraussOperators}
\end{eqnarray}
which act on the matter system.
It is a strightforward task to show that the Kraus operators indeed fulfill the defining equations $\mathcal K_{\boldsymbol n}^\dagger \mathcal K_{\boldsymbol n} \geq 0 $ and  $\sum_{\boldsymbol n}\mathcal K_{\boldsymbol n}^\dagger \mathcal K_{\boldsymbol n} = \mathbbm 1 $~\cite{Breuer2002}. 

Carrying out the multi-dimensional Fourier transformation in Eq.~\eqref{eq:kraussOperators} will be numerically very expensive in many cases. For this reason, we continue to determine the unconditional matter state using the approximated time-evolution operator in Eq.~\eqref{eq:semiclassicalTimeEvolutionOperatorApprox}, which is on an equal footing as the approximation in Eq.~\eqref{eq:momGenFkt_semiclassicalLimit}.
In doing so, we find that the reduced matter system in the semiclassical limit becomes
\begin{eqnarray}
\rho_{\text{M} }^{(\text{sc})}(t) && =\sum_{n} p_{\boldsymbol n}(t)  \rho_{\boldsymbol n}(t)   \nonumber \\
=&&   \sum_{\mu_1 , \mu_2}  \mathcal C_{\mu_1 , \mu_2   }  c_{\mu_1} c_{\mu_2} ^*    \left| u_{\mu_1,\overline {\boldsymbol \varphi}} \right>  \left<  u_{\mu_2, \overline{\boldsymbol \varphi}} \right|e^{i (E_{\mu_2, \overline{\boldsymbol \varphi} }   -E_{\mu_1,\overline{ \boldsymbol \varphi} }   )t} \nonumber,\\
\label{eq:reducedDensityMatrix_semiclassical}
\end{eqnarray}
where we have introduced the coherence integral 
\begin{equation}
\mathcal C_{\mu_1 , \mu_2   }   =  \sum_n  \sqrt{p_{\boldsymbol n  -\Delta \boldsymbol { n}_{\mu_1} (t)    }  (0)  p_{\boldsymbol n  -\Delta \boldsymbol { n}_{\mu_2} (t)    }    (0)  }
\end{equation}
as the overlap of two probability distributions, conditioned on the Floquet states $\mu_1$ and $\mu_2$, respectively. Notably, if $\mu_1=\mu_2$, the coherence integral is strictly $1$, such that  $\text{tr} \left[ \rho_{\text{M} }^{(\text{sc})}(t)\right] =1 $ for all times. Moreover, it is also easy to see that $\mathcal C_{\mu_1 , \mu_2   }\leq 1$, such that $ \rho_{\text{M} }^{(\text{sc})}(t)$ is a well-defined density matrix. We also note that Eq.~\eqref{eq:reducedDensityMatrix_semiclassical} is equivalent to the one of the PRFT investigated in Ref.~\cite{Engelhardt2024} in the semiclassical limit, for which the dynamics of the matter system has been rigorously benchmarked.

The decoherence effect is illustrated for the Jaynes-Cumming model in Fig.~\ref{fig:CountingSemiClassial.pdf}(b), which depicts the expectation value of the Pauli matrix $\hat \sigma_{\text{y}}$ as a function of time for the same three  initial states as in Fig.~\ref{fig:CountingSemiClassial.pdf}(a). The time evolution has been evaluated based on the semiclassical density  matrix in Eq.~\eqref{eq:reducedDensityMatrix_semiclassical}.

For an initial Floquet state, we observe that the expectation value is constant as a function of time. This is a special feature of the Jaynes-Cummings model in which the Floquet states are constant in time $\left|u_{\mu,\overline{ \boldsymbol \varphi}}(t) \right> = \left|u_{\mu,\overline{ \boldsymbol \varphi}}(0) \right> $ as the Hamiltonian possesses only excitation number conserving terms. More general models, e.g., the quantum Rabi model, would feature persistent oscillations which match the driving period.

For the superposition of Floquet states  depicted in the second and third column in Fig.~\ref{fig:CountingSemiClassial.pdf}(b),  we observe that the initial oscillations rapidly vanish. This is a consequence of the disappearing oscillating terms in Eq.~\eqref{eq:reducedDensityMatrix_semiclassical} which are proportional to the quickly vanishing coherence integrals $\mathcal C_{1,2} = \mathcal C_{2,1} $. The coherence integrals  decreases as the conditional mean photon flows $\Delta \boldsymbol { n}_{1} (t) =- \Delta \boldsymbol { n}_{2} (t)  $ for $\mu=1,2$ are directed in opposite directions and increase linearly in time, as can be seen in the corresponding photonic probability distribution in Fig.~\ref{fig:CountingSemiClassial.pdf}(a).

\subsection{Ambiguity in the formulation of the FCS}

According to Sec.~\ref{sec:probilities}, the probability distribution in the semiclassical limit is exclusively determined by the leading-order terms of the cumulants. Moreover, these leading order terms are generated by the minimal version of the moment-generating function in Eq.~\eqref{eq:momGenFkt_semiclassicalLimit}.  We can thus conclude that the essential information about the probability distribution is contained in the first derivatives of quasieneries with respect to the phases, i.e., $\boldsymbol \nabla_{\boldsymbol \varphi}   E_{\mu ,\overline{\boldsymbol  \varphi}}$. 

Geometrically, this means that the separation of the peaks of the probability distribution determines the leading order of the cumulants. As the distance between the peaks increases linearly with time $t$, the $l$-th cumulant $\kappa_{l}$ scales with $t^l$. This behavior is depicted for the two-mode Jaynes Cummings model in Fig.~\ref{fig:CountingSemiClassial.pdf}(c), where one can clearly observe the anticipated scaling for the initial  superposition of Floquet states. In contrast, the cumulants $\kappa_l$ for $l\ge 2$ remain approximately constant for an initial Floquet state. This behavior can be easily understood, since there is only one peak in the probability distribution for all times as can be seen in Fig.~\ref{fig:CountingSemiClassial.pdf}(a). Minor oscillations, which can be seen in some panels of Fig.~\ref{fig:CountingSemiClassial.pdf}(c), have a vanishing influence on the probability distribution as explained in Sec.~\ref{sec:probilities}.

The exclusive dependence on  $\boldsymbol \nabla_{\boldsymbol \varphi}   E_{\mu ,\overline{\boldsymbol  \varphi}}$ suggests that there is an ambiguity in the definition of the moment-generating function. This implies that different moment-generating functions will predict the same probability distribution in the  semiclassical limit, as long as their first moments for initial Floquet states agree. In Appendix~\ref{app:alternativeFCSapproaches}, we investigate this ambiguity in more detail. To do this, we compare the dynamical moment-generating function in Eq.~\eqref{eq:dynCumulantGenFct_PRFT} with the moment-generating function constructed on the basis of the photon-flux operator given in Eq.~\eqref{eq:EOF_momGenFct}, and with the moment-generating function derived in Ref.~\cite{Engelhardt2024} and given in Eq.~\eqref{eq:alt_momGenFct}. All three moment-generating functions will predict identical expressions for the first two cumulants, and the leading-order terms of the higher-order cumulants. Yet, each formalism will have different predictions for the non-leading-order terms of the  cumulants $\kappa_l$ with $l\geq 3$. Thus, according to Sec.~\ref{sec:probilities}, these alternative approaches will converge to the same semiclassical probability distribution.

This behavior is also investigated in Fig.~\ref{fig:CountingSemiClassial.pdf}(c), which shows the cumulants predicted by the three different moment-generating functions, respectively. We observe that, while the leading order coefficient agree with the accurate cumulants [corresponding to Eq.~\eqref{eq:dynCumulantGenFct_PRFT}], there are non-leading order oscillations, which gradually increase in time. Thus, while it is save to use the minimal version of the moment-generating function in Eq.~\eqref{eq:momGenFkt_semiclassicalLimit} to calculate the probability distribution in the semiclassical limit, only Eq.~\eqref{eq:dynCumulantGenFct_PRFT} will produce the accurate moments and cumulants in this regime. 

\section{Conclusions}

\label{sec:conclusions}
 
In this paper, we have investigated the scaling properties of the PRFT, a full-counting statistics approach to predict the  probability distribution of the photonic driving fields interacting with a quantum system. Based on an exact expression of the moment-generating function of the photonic probability distribution, we have discussed the scaling properties of the moments, cumulants and the probability distribution in the semiclassical limit. Interestingly, we found that the probability distribution in the semiclassical limit does not depend on the details of the moment-generating function, but only on the terms of the cumulants with leading order in time. This  allows for an ambiguity in the definition of the moment-generating function, and explains why   the PRFT in this work [Eq.~\eqref{eq:dynCumulantGenFct_PRFT}],  the  PRFT as formulated in Ref.~\cite{Engelhardt2024} [Eq.~\eqref{eq:alt_momGenFct}],  and the calculation based on the photon-flux operator [see Eq.~\eqref{eq:EOF_momGenFct}]  converge to the same probability distribution.  The  ambiguity of the moment-generating function further allowed us to derive a simple expression of the Kraus operators in the semiclassical limit, which describe the dynamics of the matter system. This enables us to simply but accurately predict the decoherence process, which is caused by the increasing light-matter entanglement. 

Given these features, we conclude that the PRFT is a flexible and accurate tool to describe the joint light-matter dynamics in the semiclassical regime. Therefore, the PRFT  will find applications in research directions where the prediction and interpretation of the photonic statistics beyond a simple mean field treatment is relevant, such as spectroscopic quantum sensing~\cite{Panahiyan2023,Dorfman2021},  generation of high-order harmonics~\cite{Jiang2024}, atomic and optical clocks~\cite{Ludlow2015,DeMille2024}, and reflectometry in semiconductor nanostructures~\cite{Vigneau2023,Cerrillo2021}. Finally, the light-matter entanglement effect discribed by the PRFT can be interpreted as a measurement process, and  thus allows for a deeper understanding of the quantum-classical decoherence process~\cite{Strasberg2022}.

\section*{Acknowledgments}

G.E. acknowledges the  support by the Guangdong Provincial Key Laboratory (Grant No.2019B121203002)
J.Y.L. acknowledges the support from the National Natural Science Foundation of China (Grant Nos. 11774311).
V.M.B. wish to thank NTT Research for their financial and technical support.
G.P. acknowledges the Spanish Ministry of Economy and Competitiveness for financial support through the grant: PID2020-117787GBI00 and  
support from CSIC Interdisciplinary Thematic Platform on Quantum  
Technologies (PTI-QTEP+).

\appendix

\section{Equivalence of the quantum and semiclassical time-evolution operator}

\label{app:quantumClassicalEquivalence}

In this Appendix, we rigorously prove the equivalence of the quantum and semiclassical time-evolution operator in Eq.~\eqref{eq:equivalenceRelationQuantumClassical}. To this end, we first introduce the basis transformation
\begin{eqnarray}
\left|\boldsymbol  n\right>  &=&  \frac{1  }{(2\pi)^{{N_{\text{D}}}/2}} \int_{-\pi}^{\pi}d\boldsymbol  \varphi    e^{-i\boldsymbol  n\cdot \boldsymbol\varphi} \left|\boldsymbol  \varphi \right>\nonumber,  \\
\left|\boldsymbol  \varphi \right>  &=&\frac{1  }{(2\pi)^{{N_{\text{D}}}/2}}  \sum_{n} e^{i\boldsymbol n \cdot \boldsymbol \varphi}\left| \boldsymbol n\right> ,
\end{eqnarray}
where the continuous multidimensinal quantum number  $\boldsymbol \varphi \in \left[ -\pi,\pi \right)^{\otimes {N_{\text{D}}}} $ takes the same role as the moment quantum number of particles in crystals. In this basis, the interaction Hamiltonian in Eq.~\eqref{eq:hamiltonian:interaction} reads
\begin{eqnarray}
\hat H^{(\text{int})} (t)  
	 &=&   \int_{-\pi}^{\pi} d\boldsymbol \varphi \left|\boldsymbol  \varphi \right> \left<\boldsymbol \varphi \right|\hat{ \mathcal H}_{\boldsymbol \varphi} (t) ,
\label{eq:hamiltonian:int}
\end{eqnarray}
where $ \hat{ \mathcal H}_{\boldsymbol \varphi} (t)$ is just the semiclassical Hamiltonian in Eq.~\eqref{eq:def:hamiltonian:semiclassical}. Thus, the basis transformation has rendered the Hamiltonian block diagonal in the $\left| \boldsymbol \varphi \right> $ basis by utilizing the translational invariance of $ \hat H^{(\text{int})} (t)  $. 

Consequently, the time-evolution operator in the interaction picture can be expressed as
\begin{eqnarray}
\hat U^{(\text{int})}(t)  &=& \hat{\mathcal T} e^{-i \int_{0}^{t} \hat H^{(\text{int})} (t') dt' } .\nonumber \\
&=& \int_{-\pi}^{\pi} d\boldsymbol \varphi \left|\boldsymbol  \varphi \right> \left<\boldsymbol \varphi \right|     \hat {\mathcal T} e^{-i \int_{0}^{t} \hat{ \mathcal H}_{\boldsymbol \varphi} (t') dt' }  .
\end{eqnarray}
Evaluating this in the Fock basis, we thus obtain
\begin{eqnarray}
\hat U^{(\text{int})}_{\boldsymbol n_{1} }(t) &=&    \left< \boldsymbol n+  \boldsymbol n_{1} \right|\hat  U^{(\text{int})}(t)\left|\boldsymbol n \right> \nonumber \\
&=&  \int_{-\pi}^{\pi}\frac{ d\boldsymbol \varphi}{(2\pi)^{N_{\text{D}}} }   e^{i \boldsymbol n_{1} \cdot \boldsymbol\varphi }    \hat{\mathcal T} e^{-i \int_{0}^{t} \hat{ \mathcal H}_{\boldsymbol \varphi} (t') dt' }  .
\end{eqnarray}
Using  finally this result to evaluate $ \hat U_{\boldsymbol \varphi}(t)$  defined in Eq.~\eqref{eq:timeEvolutionOp_fourierTrans}, we indeed find the desired relation
\begin{eqnarray}
 \hat U_{\boldsymbol \varphi}(t)   &\equiv&   \sum_{\boldsymbol n_{1}} \hat U_{\boldsymbol n_{1}}^{(\text{int})}(t) e^{-i\boldsymbol n_{1}\cdot\boldsymbol  \varphi} \nonumber \\
&=&   \hat{\mathcal T} e^{-i \int_{0}^{t} \hat{ \mathcal H}_{\boldsymbol \varphi} (t') dt' } \nonumber \\
 &=&  \hat {\mathcal U}_{\boldsymbol \varphi}(t) ,
\end{eqnarray}
where in the last line we have identified the definition of the semiclassical time-evolution operator in terms of the time-ordering operators and the semiclassical Hamitonian given in Eq.~\eqref{eq:semiclassicalTimeEvolutionOperator}.

\section{Error analysis}

\label{app:errorAnalysis}

Here, we carry out a detailed error analysis for the cumulants with respect to the semiclassical approximation in Eq.~\eqref{eq:dynCumulantGenFct_PRFT}.

Before delving into the details of the error analysis, we first introduce the concept of dynamical moments and cumulants. The dynamical moment-generating function has been  defined in Eq.~\eqref{eq:dynCumulantGenFct}. Accordingly, we define the dynamical cumulant-generating function via
\begin{equation}
	K_{\text{dy} ,\boldsymbol \chi}(t)  = \log M_{\text{dy} ,\boldsymbol \chi}(t) .
\end{equation}
The dynamical moments and cumulants are defined as the derivatives with respect to the counting fields, i.e.,
\begin{eqnarray}
m_{ \text{dy} ,l}^{(k)} &=&  \frac{d^l}{d\,(-i\chi_k)^l} M_{\text{dy} ,\boldsymbol \chi= \boldsymbol 0}  ,\nonumber \\
\label{eq:def:dynCumulants}
\kappa_{ \text{dy} ,l}^{(k)} &=& \frac{d^l}{d\,(-i\chi_k)^l} K_{\text{dy} ,\boldsymbol \chi= \boldsymbol 0}  .
\end{eqnarray}
As the moment-generating function factorizes according to  Eq.~\eqref{eq:momentGenFkt_factorized}, the dynamical cumulant-generating function is simply a sum of the cumulant-generating function at time $t=0$ and the dynamical cumulant-generating function, i.e., $K_{\boldsymbol \chi} (t)  =K_{\text{dy} ,\boldsymbol \chi} (t)  +K_{\boldsymbol \chi} (0) $. Consequently, the cumulants fulfill 
\begin{equation}
\kappa_{ l}^{(k)} (t) = \kappa_{ \text{dy} ,l}^{(k)}(t)  + \kappa_{l}^{(k)} (0) .
\label{eq:cumulant_timeEvolution}
\end{equation}
Moreover,  dynamical cumulants $  \kappa _{\text{dy},1}  $ can be obtained from the dynamical moments via the relations
\begin{eqnarray}
\kappa _{\text{dy},1}   &=&    m_{\text{dy},1 } ,\nonumber  \\
\kappa _{\text{dy},2}   &=&   m_{\text{dy},2  }   -  m_{\text{dy},1  }^2 ,\nonumber   \\  
\kappa _{\text{dy},3}   &=&   m_{\text{dy},3  } - 3m_{\text{dy},2  }m_{\text{dy},1  }  +2 m_{\text{dy},1  } ^3 , \nonumber \\  
\cdots
\label{eq:cumulant_moment_relation}
\end{eqnarray}	
and similar for higher-order cumulants. These relations for the dynamical cumulants are thus equivalent to the relations for the common cumulants, and can be proven in the same way. From Eq.~\eqref{eq:cumulant_timeEvolution} and Eq.~\eqref{eq:cumulant_moment_relation} we thus see that we can infer the (error) scaling of the cumulants by analyzing the scaling properties of the dynamical moments.

In the following, we discuss the scaling properties of the dynamical  moments  based on the Floquet expansion in Eq.~\eqref{eq:floquetExpansion}.  To keep the notation simple, we focus here on a single mode $k=1$ with mean phase $\overline \varphi=0$, and variance $ \sigma^2$.  Generalization to the multimode case is straightforward.
Differentiating the exact expression of the dynamical moment-generating function in Eq.~\eqref{eq:dynCumulantGenFct}  $l$ times with respect to the counting field $\chi_1$, we identify the  leading term of the $l$-th dynamical moment to be
  \begin{eqnarray}
 m_{\text{dy} ,l}  (t)
 &=&\sum_{\nu,\mu,\mu^\prime}  c_{\mu}^{*} c_{\mu^\prime}     \nonumber \\
 &&\times  \frac{\sigma}{\sqrt{\pi} } \int  d\varphi     \left( E_{\nu,  \varphi } ^\prime  t  \right)^l     
 e^{-    \sigma^2   \varphi^{ 2}  } \nonumber  \\
  &&\quad\times\left<  u_{\mu, \overline{\varphi}   } \mid u_{\nu, \varphi } \right> \left<  u_{\nu, \varphi } \mid u_{\mu^\prime, \overline{\varphi}   }  \right>   \nonumber  \\
 &&+\mathcal O\left( t^{l-1}\right),
 \label{eq:kth-moment_floquetExpansion} 
 \end{eqnarray}
 where $E_{\nu, \varphi  } ^\prime =  dE_{\mu,\varphi }/d\varphi$. The non-leading order terms in time $\mathcal O\left( t^{l-1}\right)$ originate form high-order derivatives of the quasienergy $d^{r}E_{\mu,\varphi }/d\varphi^r$ and derivatives of the counting-field-dependend Floquet states. To make progress, we expand $E_{\nu, \varphi } ^\prime $ in a Taylor series up to second order in $\varphi$ around $\overline{\varphi}=0$,
\begin{eqnarray}
&&E_{\nu,  \varphi } ^\prime   =  E_{\nu,\overline{\varphi}}^\prime  +  E_{\nu,\overline{\varphi} }^{\prime\prime }\varphi +  \frac{1}{2}E_{\nu,\overline{\varphi} }^{\prime\prime \prime } \varphi^{2 } ,\nonumber\\ 
 &&\left<  u_{\mu, \overline{\varphi}   } \mid u_{\nu, \varphi } \right> \left<  u_{\nu, \varphi } \mid u_{\mu^\prime, \overline{\varphi}   }  \right>  \nonumber \\
 & &= \delta_{\mu,\nu} \delta_{\mu,\mu^\prime}  +  g_{\mu,\nu,\mu^\prime}^{(1)}\varphi +  \frac{1}{2} g_{\mu,\nu,\mu^\prime}^{(2)}\varphi^{ 2 },
\end{eqnarray}
where $E_{\nu,\overline{\varphi}  }^{\prime\prime } $  and  $E_{\nu,\overline{\varphi} }^{\prime\prime \prime } $ denote the second and third derivatives of the quasienergy with respect to $\varphi$ evaluated at $\overline \varphi = 0$. Likewise, $g_{\mu,\nu,\mu^\prime }^{(j)}$ are the Taylor expansion coefficients of the  Floquet state scalar products. Inserting these expansions into Eq.~\eqref{eq:kth-moment_floquetExpansion}  and evaluating the Gaussian integrals, we find
  \begin{eqnarray}
m_{\text{dy} ,l}  
&&=t^l\sum_{\nu,\mu,\mu^\prime } c_{\mu}^* c_{\nu} \left\lbrace  \nonumber \delta_{\mu,\nu} \delta_{\mu,\mu^\prime}  \left(  E_{\nu,\overline{\varphi} }^\prime   \right)^l      \right.    \\
&&+ \frac{1}{\sigma^2} \left[ l \left(  E_{\nu,\overline{\varphi} }^\prime   \right)^{l-1}    E_{\nu,\overline{ \varphi}}^{\prime\prime\prime}  \delta_{\mu,\nu}, \delta_{\mu,\mu^\prime}   \right.\nonumber  \\
&& + l(l-1)\left(  E_{\nu,\overline{ \varphi} }^\prime   \right)^{l-2}    \left( E_{\nu,\overline{ \varphi} }^{\prime \prime} \right)^2  \delta_{\mu,\nu} \delta_{\mu,\mu^\prime}   \nonumber \\
&&  +  l\left(  E_{\nu,\overline{\varphi} }^\prime   \right)^{l-1}    E_{\nu,\overline{ \varphi} }^{\prime\prime}         g_{\mu,\nu,\mu^\prime}^{(1)}   +   \left.\left.  \left(  E_{\nu,\overline{\varphi} }^\prime   \right)^l g_{\mu,\nu,\mu^\prime}^{(2)}  \right]  \right\rbrace \nonumber \\
 && +\mathcal O \left( \frac{t^l}{\sigma^4},  t^{l-1}  \right) .
\end{eqnarray}
The first line contains the leading-order terms in time  in the thermodynamic limit $\sigma\rightarrow \infty$. Interestingly, they are exclusively determined by the first derivative of the quasienergies $ E_{\mu,\overline{ \varphi}}^\prime $. 	The second till fourth lines scale also with $\propto t^l$, yet, they are renormalized with $1/\sigma^2$, such that for a fixed time they will vanish in the semiclassical limit $\sigma \rightarrow \infty$. The terms which remain in this limit are identical to the moments  generated by the moment-generating function in the PRFT in Eq.~\eqref{eq:dynCumulantGenFct_PRFT}. For this reason, we conclude that the exact moments, $ m_{\text{dy} ,l} (t)$, are related  to the moments predicted by the PRFT $ m_{\text{dy} ,l}^{(\text{PRFT})} (t)$ via the scaling relation
\begin{eqnarray}
m_{\text{dy} ,l} (t)   = m_{\text{dy} ,l}^{(\text{PRFT})} (t) +  \mathcal O \left( \frac{t^l}{\sigma^2}   \right) .
\label{eq:d} 
\end{eqnarray}
 From Eq.~\eqref{eq:cumulant_timeEvolution} and Eq.~\eqref{eq:cumulant_moment_relation}, we can directly infer that the cumulants share the same scaling properties as the moments, namely
\begin{eqnarray}
\kappa_l (t)   = \kappa_l^{(\text{PRFT})} (t) +  \mathcal O \left( \frac{t^l}{\sigma^2}   \right) ,
 \label{eq:d} 
\end{eqnarray}
 where  $ \kappa_l^{(\text{PRFT})} (t) $ denote the  cumulants predicted by the PRFT.

\section{Alternative FCS approaches}

\label{app:alternativeFCSapproaches}

As discussed in Sec.~\ref{sec:probilities}, the semiclassical probability distribution is determined by the leading-order terms in time of the cumulants. This conclusion suggests some arbitrariness in the definition of the moment-generating function. Here we discuss two alternative approaches to calculate the photonic probability distribution.

The most widely used approach to gain information about the state of the photonic system is in terms of the photon-flux operator, which is commonly defined as
\begin{equation}
\hat j_k (t) =\frac{d}{d\varphi_k} \hat {\mathcal H}_{ \boldsymbol\varphi = \overline{\boldsymbol\varphi} }(t) 
\end{equation}
in the Schroedinger picture, and as
\begin{equation}
\tilde  j_k (t) =\hat{\mathcal U}_{ \overline{\boldsymbol\varphi}   }^{\dagger} (t)  \hat j_k (t) \hat{\mathcal U}_{ \overline{\boldsymbol\varphi}  } (t)
\end{equation}
in the Heisenberg picture. In this spirit, one can motivate the dynamical moment-generating function as the matrix exponential of the time-integrated photon-flux operator
\begin{equation}
M_{\text{dy},\boldsymbol \chi}^{(\text{PFO})}(t)  =    \left<      e^{-i \sum_k  \chi_k \int_{0}^{t}  \tilde  j_k (t)   } \right>_{t=0}.
\label{eq:EOF_momGenFct}
\end{equation}
The corresponding moments and cumulants are multi-time correlation functions of the photon-flux operator (see Sec.~\ref{app:lowOrderCumulants}).  

The alternative approach of the PRFT in Ref.~\cite{Engelhardt2024}  [which will be called alternative PRFT (aPRFT) in the following] derives the following version of the dynamical moment-generating function
\begin{equation}
M_{\text{dy},\boldsymbol \chi}^{(\text{aPRFT})}( t)  =   \frac{1}{2} \left<    \hat { \mathcal U}_{ \overline{\boldsymbol\varphi} }^{\dagger} (t)  \hat {\mathcal U}_{\overline{\boldsymbol\varphi} +\boldsymbol \chi}(t)  +     \hat{\mathcal U}_{\overline{\boldsymbol\varphi}  -\boldsymbol \chi}^{\dagger} (t)  \hat{\mathcal U}_{\overline{\boldsymbol\varphi} }(t)   \right>_{t=0} .
\label{eq:alt_momGenFct}
\end{equation}
In the following, we will compare the alternative versions of the moment-generating function in Eq.~\eqref{eq:EOF_momGenFct} and Eq.~\eqref{eq:alt_momGenFct}   with the PRFT introduced in Eq.~\eqref{eq:dynCumulantGenFct_PRFT}. In doing so, we will show that, while all three dynamical moment-generating functions will predict the same probability distribution in the semiclassical limit, they will make distinct predictions for the non-leading order terms in time of higher-order cumulants $\kappa_l$ with $l\geq3$.

\subsection{Low-order cumulants}

\label{app:lowOrderCumulants}

To compare the low-order cumulants of the three different versions of the dynamical-moment generating functions, we first consider the first two  dynamical moments of the photon-flux operator method in Eq.~\eqref{eq:EOF_momGenFct}. Carrying out the derivatives with respect to the counting field, the first moment reads as
\begin{equation}
m_{\text{dy}, 1}^{(k,\text{PFO})} =  \left< 	\int_{0}^{t} dt_1  \tilde  j_k (t_1) \right>_{t=0},
\label{eq:firstMoment_EOF}
\end{equation}
while the second moment becomes
\begin{equation}
m_{\text{dy}, 2}^{(k,\text{PFO})} =  	\int_{0}^{t}  dt_1 	\int_{0}^{t}  dt_2 	 \left<  \tilde  j_k (t_1) \tilde  j_k (t_2)  \right>_{t=0} 
\label{eq:secondMoment_EOF}.
\end{equation}
Interestingly, these two expressions agree with the first two dynamical moments as predicted by   Eq.\eqref{eq:dynCumulantGenFct_PRFT} and Eq.\eqref{eq:alt_momGenFct} as we will show immediately. Consequently, the first two cumulants are equal for the three semiclassical moment-generating functions.

The third moment predicted by the photon-flux operator reads as
\begin{equation}
m_{\text{dy}, 3}^{(k,\text{PFO})} =  	\int_{0}^{t} dt_1 	\int_{0}^{t} dt_2  \int_{0}^{t}  dt_3	 \left<  \tilde  j_k (t_1) \tilde  j_k (t_2)  \tilde  j_k (t_3)  \right>_{t=0}   .
\label{eq:thirdMoment_EOF}
\end{equation}
However, neither version of the PRFT can reproduce this form. As we explain in the following, the PRFT can only generate correlations functions, which  have a particular order in time. In  contrast, the operators in Eq.~\eqref{eq:thirdMoment_EOF} can have an arbitrary time order.
 
To investigate the time order generated by the PRFT, we first distribute the semiclassical Hamiltonian in Eq.~\eqref{eq:def:hamiltonian:semiclassical} as follows
\begin{eqnarray}
 \hat {\mathcal H}_{\overline{\boldsymbol\varphi} +\boldsymbol  \chi}  = \hat{\mathcal H}_{\overline{\boldsymbol\varphi}} (t) + \hat V_{\boldsymbol  \chi}(t),
\end{eqnarray}
where the first operator is defined in Eq.~\eqref{eq:def:hamiltonian:semiclassical}, and the second operator is given by
\begin{equation}
\hat V_{\boldsymbol  \chi}(t)  = \sum_k 2g \alpha_k  \hat  V_k  \left[   \cos (\omega_k t - \overline{\varphi}_k -\chi_k)   - \cos (\omega_k t -\overline{ \varphi}_k)  \right] .
\end{equation}
Now we transform the system into an interaction picture defined by $\hat{\mathcal U}_{\overline{\boldsymbol\varphi}}(t)  = \hat{\mathcal T} \exp \left[ -i \int_{0}^{t} \hat {\mathcal H}_{\overline{\boldsymbol\varphi}} (t^\prime)dt^\prime  \right]$, such that the time-evolution operator in the lab frame factorizes as
\begin{equation}
\hat {\mathcal U}_{\overline{\boldsymbol\varphi}+\boldsymbol \chi}(t) = \hat{\mathcal U}_{\overline{\boldsymbol\varphi}}(t) \tilde {\mathcal U}_{\overline{\boldsymbol\varphi}, \boldsymbol \chi} (t),
\end{equation}
where  time evolution operator in the interaction picture can be expanded as
\begin{eqnarray}
\tilde {\mathcal U}_{\overline{\boldsymbol\varphi}, \boldsymbol \chi} &=&\hat {\mathcal T}  e^{-i \int_{0}^{t}   \hat V_ {\boldsymbol \chi}(t^\prime) dt^\prime    } \nonumber \\
&=&  \mathbbm 1 -i \int_{0}^{t} dt_1  \hat V_{\boldsymbol  \chi}(t_1)  -   \int_{0}^{t} dt_1  \int_{0}^{t_1} dt_2 \hat  V_{\boldsymbol  \chi}(t_1) \hat   V_{\boldsymbol  \chi}(t_2)  \nonumber  \\
&&+i     \int_{0}^{t} dt_1  \int_{0}^{t_1} dt_2 \int_{0}^{t_2} dt_3  \hat V_{\boldsymbol  \chi}(t_1)\hat    V_{\boldsymbol  \chi}(t_2) \hat 	V_{\boldsymbol  \chi}(t_3) \nonumber  \\
	 &&+\cdots
	 \label{eq:timeEvolutionOperator_interactionPicture}
\end{eqnarray}
Crucially, the perturbation operators $\hat V_{\boldsymbol  \chi}(t)  $ are strictly time-ordered within each perturbation order.

In the interaction picture, the dynamical moment-generating function of the PRFT reads
\begin{equation}
M_{\text{dy},\boldsymbol \chi}^{(\text{PRFT})}( t)  =    \left<     \tilde  {\mathcal U}_{ \overline{\boldsymbol\varphi} ,-\boldsymbol \chi/2 }^{\dagger} (t)  \tilde {\mathcal U}_{\overline{\boldsymbol\varphi}, \boldsymbol \chi/2}(t)  \right>_{t=0}.
\label{eq:dynMomGenFkt_interactionPicture}
\end{equation}
When inserting the expansion in Eq.~\eqref{eq:timeEvolutionOperator_interactionPicture} into Eq.~\eqref{eq:dynMomGenFkt_interactionPicture}, we can evaluate the dynamical moments by deriving with respect to the counting fields $\chi_k$. Thereby, we can take advantage of the fact that $\tilde  V_{\boldsymbol  \chi = \boldsymbol 0}(t)=0$  and
\begin{equation}
\frac{d}{d\chi_k} \tilde  V_{\boldsymbol  \chi = \boldsymbol 0}(t)  =  \tilde  j_k (t) .
\end{equation}
In doing so, we find that the first two dynamical moments indeed become Eq.~\eqref{eq:firstMoment_EOF} and Eq.~\eqref{eq:secondMoment_EOF}, respectively. 

However, the third dynamical cumulant in Eq.~\eqref{eq:thirdMoment_EOF} contains terms like $\tilde  j_k (t_1)\tilde  j_k (t_2)\tilde  j_k (t_3)$ with $t_2<t_1$ and $t_2<t_3$. Careful inspection of Eq.~\eqref{eq:timeEvolutionOperator_interactionPicture} reveals that such terms cannot be produced by the moment-generating function in  Eq.~\eqref{eq:dynMomGenFkt_interactionPicture} because of the strict time order of the operators $\tilde  V_{\boldsymbol  \chi }(t)$. The same arguments hold also for the moment-generating function of the aPRFT in Eq.~\eqref{eq:alt_momGenFct}.

\subsection{Scaling analysis of different counting approaches}

Here, we will show that the two moment-generating functions in Eq.~\eqref{eq:EOF_momGenFct} and Eq.~\eqref{eq:alt_momGenFct} share the same leading-order terms in time as the moment-generating of the photon-resolved Floquet theory in Eq.\eqref{eq:dynCumulantGenFct_PRFT}. For the moment-generating function of the PRFT in Eq.~\eqref{eq:dynCumulantGenFct_PRFT} and  the moment-generating function of the aPRFT in Eq.~\eqref{eq:alt_momGenFct}, it is straightforward to show that the leading orders  in time of the dynamical moments are given by
  \begin{eqnarray}
m_{\text{dy},l}^{(k,\text{PRFT} )} &=& m_{\text{dy},l}^{(k,\text{aPRFT})}+ \mathcal O(t^{l-1})\nonumber \\
&=&t^l\sum_{\mu } \left| c_{\mu}\right|^2   \left( \frac{d}{d\varphi_k}  E_{\mu,\boldsymbol {\overline \varphi}}   \right)^l  + \mathcal O(t^{l-1})      \nonumber \\
\end{eqnarray}
by using the Floquet expansion of the respective dynamical-moment-generating functions.

On the other hand, the dynamical $l$-th moment obtained by the moment-generating function in Eq.~\eqref{eq:EOF_momGenFct}  can be expressed in the Floquet basis as
\begin{eqnarray}
m_{\text{dy},l}^{(k, \text{PFO})} &=&  \sum_{\mu,\nu} c_\mu^* c_\nu  \left<  u_{\mu}	 \right|\left[ \int_{0}^{t}  dt_1	 \tilde  j_k (t_1)\right]^l \left| u_{\nu}  \right>_{t=0}   \nonumber \\
&=&  t^l   \sum_{\mu} \left| c_\mu \right|^2  \left(  \frac{1}{\tau} \int_{0}^{\tau}  dt \left<  u_{\mu}(t)	 \right|     \hat  j_k (t)  \left|  u_{\mu}(t)	 \right>   \right)^l \nonumber  \\
&&+ \mathcal O(t^{l-1})\nonumber \\
&=&    t^l \sum_{\mu} \left| c_\mu \right|^2  \left(  \frac{d}{d\varphi_k} E_{\mu, \overline{\boldsymbol\varphi} }  \right)^l  + \mathcal O(t^{l-1}).
\end{eqnarray}
In the second equality, we have identified the leading-order terms in time, which is given  by the diagonal elements (in the Floquet basis) of the time-averaged photon-flux operator $ \hat  j_k (t)$. In the third equality, we took advantage of the Hellman-Feyman theorem for Floquet systems.
As the leading-order terms in time of the three dynamical moments agree for all approaches, this is also valid for the leading-order terms of the  cumulants because of the relations in Eq.~\eqref{eq:cumulant_moment_relation}. 

For completeness, we finally prove the Hellman-Feynman theorem in periodically-driven systems, which are characterized by a time-periodic Hamiltonian $\hat{\mathcal H}(t) =\hat {\mathcal H}(t+\tau) $. According to Floquet theory, we can expand the time-evolution operator as $\hat {\mathcal U}  = \sum_\mu e^{-iE_{\mu }t }\left| u_\mu(t) \right> \left<u_\mu(0)\right| $ with the quasienergies $E_{\mu }$ and the $\tau$ periodic Floquet states $\left| u_\mu(t) \right> $. From Floquet theory~\cite{Shirley1965} it is known that
\begin{equation}
	\left [ \hat{\mathcal H}(t)-i\frac{d}{dt}  \right] \left| u_\mu(t) \right> = E_{\mu }\left| u_\mu(t) \right>,
\end{equation} 
where the operator on the left-hand side is regarded as a operator in space-time coordinates.

The Hellman-Feynman theorem  states  that, if $\hat A(t) = \frac{d}{d\lambda}\mathcal H(t)$, then the time-averaged expectation value in a Floquet state can be expressed by a derivative of the quasienergy with respect to $\lambda$. More precisely,
\begin{eqnarray}
	\overline{ \left< \hat A (t)\right> } &\equiv&\frac{1}{\tau} \int_{0}^{\tau}  \left<  u_{\mu}(t)	 \right|     \hat  A (t)  \left|  u_{\mu}(t)	 \right>  dt \nonumber \\
	 &=& \frac{1}{\tau} \int_{0}^{\tau}  \left<  u_{\mu}(t)	 \right|    \frac{d}{d\lambda}\hat {\mathcal H}(t)    \left|  u_{\mu}(t)	 \right>  dt \nonumber \\
	 &=& \frac{1}{\tau} \int_{0}^{\tau}  \left<  u_{\mu}(t)	 \right|  \frac{d}{d\lambda}  \left [ \hat {\mathcal H}(t)-i\frac{d}{dt}  \right]   \left|  u_{\mu}(t)	 \right>  dt\nonumber  \\
	  &=&\frac{d}{d\lambda} \frac{1}{\tau} \int_{0}^{\tau}  \left<  u_{\mu}(t)	 \right|    \left [ \hat {\mathcal H}(t)-i\frac{d}{dt}  \right]   \left|  u_{\mu}(t)	 \right>  dt  \nonumber \\
	  &=& \frac{d}{d\lambda}  \frac{1}{\tau} \int_{0}^{\tau}  E_{\mu}  dt = \frac{d}{d\lambda} E_{\mu} ,
\end{eqnarray}
proving thus the Hellman-Feynman theorem.

 \bibliography{mybibliography}

\begin{thebibliography}{66}%
\makeatletter
\providecommand \@ifxundefined [1]{%
 \@ifx{#1\undefined}
}%
\providecommand \@ifnum [1]{%
 \ifnum #1\expandafter \@firstoftwo
 \else \expandafter \@secondoftwo
 \fi
}%
\providecommand \@ifx [1]{%
 \ifx #1\expandafter \@firstoftwo
 \else \expandafter \@secondoftwo
 \fi
}%
\providecommand \natexlab [1]{#1}%
\providecommand \enquote  [1]{``#1''}%
\providecommand \bibnamefont  [1]{#1}%
\providecommand \bibfnamefont [1]{#1}%
\providecommand \citenamefont [1]{#1}%
\providecommand \href@noop [0]{\@secondoftwo}%
\providecommand \href [0]{\begingroup \@sanitize@url \@href}%
\providecommand \@href[1]{\@@startlink{#1}\@@href}%
\providecommand \@@href[1]{\endgroup#1\@@endlink}%
\providecommand \@sanitize@url [0]{\catcode `\\12\catcode `\$12\catcode
  `\&12\catcode `\#12\catcode `\^12\catcode `\_12\catcode `\%12\relax}%
\providecommand \@@startlink[1]{}%
\providecommand \@@endlink[0]{}%
\providecommand \url  [0]{\begingroup\@sanitize@url \@url }%
\providecommand \@url [1]{\endgroup\@href {#1}{\urlprefix }}%
\providecommand \urlprefix  [0]{URL }%
\providecommand \Eprint [0]{\href }%
\providecommand \doibase [0]{https://doi.org/}%
\providecommand \selectlanguage [0]{\@gobble}%
\providecommand \bibinfo  [0]{\@secondoftwo}%
\providecommand \bibfield  [0]{\@secondoftwo}%
\providecommand \translation [1]{[#1]}%
\providecommand \BibitemOpen [0]{}%
\providecommand \bibitemStop [0]{}%
\providecommand \bibitemNoStop [0]{.\EOS\space}%
\providecommand \EOS [0]{\spacefactor3000\relax}%
\providecommand \BibitemShut  [1]{\csname bibitem#1\endcsname}%
\let\auto@bib@innerbib\@empty
\bibitem [{\citenamefont {Bastidas}\ \emph {et~al.}(2012)\citenamefont
  {Bastidas}, \citenamefont {Emary}, \citenamefont {Regler},\ and\
  \citenamefont {Brandes}}]{Bastidas2012}%
  \BibitemOpen
  \bibfield  {author} {\bibinfo {author} {\bibfnamefont {V.~M.}\ \bibnamefont
  {Bastidas}}, \bibinfo {author} {\bibfnamefont {C.}~\bibnamefont {Emary}},
  \bibinfo {author} {\bibfnamefont {B.}~\bibnamefont {Regler}},\ and\ \bibinfo
  {author} {\bibfnamefont {T.}~\bibnamefont {Brandes}},\ }\bibfield  {title}
  {\bibinfo {title} {Nonequilibrium quantum phase transitions in the {D}icke
  model},\ }\href {https://doi.org/10.1103/PhysRevLett.108.043003} {\bibfield
  {journal} {\bibinfo  {journal} {Phys. Rev. Lett.}\ }\textbf {\bibinfo
  {volume} {108}},\ \bibinfo {pages} {043003} (\bibinfo {year}
  {2012})}\BibitemShut {NoStop}%
\bibitem [{\citenamefont {Engelhardt}\ \emph {et~al.}(2013)\citenamefont
  {Engelhardt}, \citenamefont {Bastidas}, \citenamefont {Emary},\ and\
  \citenamefont {Brandes}}]{Engelhardt2013}%
  \BibitemOpen
  \bibfield  {author} {\bibinfo {author} {\bibfnamefont {G.}~\bibnamefont
  {Engelhardt}}, \bibinfo {author} {\bibfnamefont {V.~M.}\ \bibnamefont
  {Bastidas}}, \bibinfo {author} {\bibfnamefont {C.}~\bibnamefont {Emary}},\
  and\ \bibinfo {author} {\bibfnamefont {T.}~\bibnamefont {Brandes}},\
  }\bibfield  {title} {\bibinfo {title} {{ac-driven quantum phase transition in
  the Lipkin-Meshkov-Glick model}},\ }\href
  {https://doi.org/10.1103/PhysRevE.87.052110} {\bibfield  {journal} {\bibinfo
  {journal} {Phys. Rev. E}\ }\textbf {\bibinfo {volume} {87}},\ \bibinfo
  {pages} {052110} (\bibinfo {year} {2013})}\BibitemShut {NoStop}%
\bibitem [{\citenamefont {Rudner}\ \emph {et~al.}(2013)\citenamefont {Rudner},
  \citenamefont {Lindner}, \citenamefont {Berg},\ and\ \citenamefont
  {Levin}}]{rudner2013anomalous}%
  \BibitemOpen
  \bibfield  {author} {\bibinfo {author} {\bibfnamefont {M.~S.}\ \bibnamefont
  {Rudner}}, \bibinfo {author} {\bibfnamefont {N.~H.}\ \bibnamefont {Lindner}},
  \bibinfo {author} {\bibfnamefont {E.}~\bibnamefont {Berg}},\ and\ \bibinfo
  {author} {\bibfnamefont {M.}~\bibnamefont {Levin}},\ }\bibfield  {title}
  {\bibinfo {title} {Anomalous edge states and the bulk-edge correspondence for
  periodically driven two-dimensional systems},\ }\href@noop {} {\bibfield
  {journal} {\bibinfo  {journal} {Phys. Rev. X}\ }\textbf {\bibinfo {volume}
  {3}},\ \bibinfo {pages} {031005} (\bibinfo {year} {2013})}\BibitemShut
  {NoStop}%
\bibitem [{\citenamefont {Benito}\ \emph {et~al.}(2014)\citenamefont {Benito},
  \citenamefont {G\'omez-Le\'on}, \citenamefont {Bastidas}, \citenamefont
  {Brandes},\ and\ \citenamefont {Platero}}]{Benito2014}%
  \BibitemOpen
  \bibfield  {author} {\bibinfo {author} {\bibfnamefont {M.}~\bibnamefont
  {Benito}}, \bibinfo {author} {\bibfnamefont {A.}~\bibnamefont
  {G\'omez-Le\'on}}, \bibinfo {author} {\bibfnamefont {V.~M.}\ \bibnamefont
  {Bastidas}}, \bibinfo {author} {\bibfnamefont {T.}~\bibnamefont {Brandes}},\
  and\ \bibinfo {author} {\bibfnamefont {G.}~\bibnamefont {Platero}},\
  }\bibfield  {title} {\bibinfo {title} {Floquet engineering of long-range
  $p$-wave superconductivity},\ }\href
  {https://doi.org/10.1103/PhysRevB.90.205127} {\bibfield  {journal} {\bibinfo
  {journal} {Phys. Rev. B}\ }\textbf {\bibinfo {volume} {90}},\ \bibinfo
  {pages} {205127} (\bibinfo {year} {2014})}\BibitemShut {NoStop}%
\bibitem [{\citenamefont {Engelhardt}\ \emph {et~al.}(2016)\citenamefont
  {Engelhardt}, \citenamefont {Benito}, \citenamefont {Platero},\ and\
  \citenamefont {Brandes}}]{Engelhardt2016}%
  \BibitemOpen
  \bibfield  {author} {\bibinfo {author} {\bibfnamefont {G.}~\bibnamefont
  {Engelhardt}}, \bibinfo {author} {\bibfnamefont {M.}~\bibnamefont {Benito}},
  \bibinfo {author} {\bibfnamefont {G.}~\bibnamefont {Platero}},\ and\ \bibinfo
  {author} {\bibfnamefont {T.}~\bibnamefont {Brandes}},\ }\bibfield  {title}
  {\bibinfo {title} {Topological instabilities in ac-driven bosonic systems},\
  }\href {https://doi.org/10.1103/PhysRevLett.117.045302} {\bibfield  {journal}
  {\bibinfo  {journal} {Phys. Rev. Lett.}\ }\textbf {\bibinfo {volume} {117}},\
  \bibinfo {pages} {045302} (\bibinfo {year} {2016})}\BibitemShut {NoStop}%
\bibitem [{\citenamefont {Roy}\ and\ \citenamefont
  {Harper}(2017)}]{roy2017floquet}%
  \BibitemOpen
  \bibfield  {author} {\bibinfo {author} {\bibfnamefont {R.}~\bibnamefont
  {Roy}}\ and\ \bibinfo {author} {\bibfnamefont {F.}~\bibnamefont {Harper}},\
  }\bibfield  {title} {\bibinfo {title} {Floquet topological phases with
  symmetry in all dimensions},\ }\href@noop {} {\bibfield  {journal} {\bibinfo
  {journal} {Phys. Rev. B}\ }\textbf {\bibinfo {volume} {95}},\ \bibinfo
  {pages} {195128} (\bibinfo {year} {2017})}\BibitemShut {NoStop}%
\bibitem [{\citenamefont {Maczewsky}\ \emph {et~al.}(2017)\citenamefont
  {Maczewsky}, \citenamefont {Zeuner}, \citenamefont {Nolte},\ and\
  \citenamefont {Szameit}}]{maczewsky2017observation}%
  \BibitemOpen
  \bibfield  {author} {\bibinfo {author} {\bibfnamefont {L.~J.}\ \bibnamefont
  {Maczewsky}}, \bibinfo {author} {\bibfnamefont {J.~M.}\ \bibnamefont
  {Zeuner}}, \bibinfo {author} {\bibfnamefont {S.}~\bibnamefont {Nolte}},\ and\
  \bibinfo {author} {\bibfnamefont {A.}~\bibnamefont {Szameit}},\ }\bibfield
  {title} {\bibinfo {title} {{Observation of photonic anomalous Floquet
  topological insulators}},\ }\href@noop {} {\bibfield  {journal} {\bibinfo
  {journal} {Nature communications}\ }\textbf {\bibinfo {volume} {8}},\
  \bibinfo {pages} {1} (\bibinfo {year} {2017})}\BibitemShut {NoStop}%
\bibitem [{\citenamefont {Schnell}\ \emph {et~al.}(2023)\citenamefont
  {Schnell}, \citenamefont {Weitenberg},\ and\ \citenamefont
  {Eckardt}}]{Schnell2023}%
  \BibitemOpen
  \bibfield  {author} {\bibinfo {author} {\bibfnamefont {A.}~\bibnamefont
  {Schnell}}, \bibinfo {author} {\bibfnamefont {C.}~\bibnamefont
  {Weitenberg}},\ and\ \bibinfo {author} {\bibfnamefont {A.}~\bibnamefont
  {Eckardt}},\ }\href@noop {} {\bibinfo {title} {Dissipative preparation of a
  floquet topological insulator in an optical lattice via bath engineering}}
  (\bibinfo {year} {2023}),\ \Eprint {https://arxiv.org/abs/2307.03739}
  {arXiv:2307.03739 [cond-mat.quant-gas]} \BibitemShut {NoStop}%
\bibitem [{\citenamefont {Khemani}\ \emph {et~al.}(2016)\citenamefont
  {Khemani}, \citenamefont {Lazarides}, \citenamefont {Moessner},\ and\
  \citenamefont {Sondhi}}]{khemani2016phase}%
  \BibitemOpen
  \bibfield  {author} {\bibinfo {author} {\bibfnamefont {V.}~\bibnamefont
  {Khemani}}, \bibinfo {author} {\bibfnamefont {A.}~\bibnamefont {Lazarides}},
  \bibinfo {author} {\bibfnamefont {R.}~\bibnamefont {Moessner}},\ and\
  \bibinfo {author} {\bibfnamefont {S.~L.}\ \bibnamefont {Sondhi}},\ }\bibfield
   {title} {\bibinfo {title} {Phase structure of driven quantum systems},\
  }\href {https://doi.org/10.1103/PhysRevLett.116.250401} {\bibfield  {journal}
  {\bibinfo  {journal} {Phys. Rev. Lett.}\ }\textbf {\bibinfo {volume} {116}},\
  \bibinfo {pages} {250401} (\bibinfo {year} {2016})}\BibitemShut {NoStop}%
\bibitem [{\citenamefont {Yao}\ \emph {et~al.}(2017)\citenamefont {Yao},
  \citenamefont {Potter}, \citenamefont {Potirniche},\ and\ \citenamefont
  {Vishwanath}}]{yao2017discrete}%
  \BibitemOpen
  \bibfield  {author} {\bibinfo {author} {\bibfnamefont {N.~Y.}\ \bibnamefont
  {Yao}}, \bibinfo {author} {\bibfnamefont {A.~C.}\ \bibnamefont {Potter}},
  \bibinfo {author} {\bibfnamefont {I.-D.}\ \bibnamefont {Potirniche}},\ and\
  \bibinfo {author} {\bibfnamefont {A.}~\bibnamefont {Vishwanath}},\ }\bibfield
   {title} {\bibinfo {title} {Discrete time crystals: Rigidity, criticality,
  and realizations},\ }\href {https://doi.org/10.1103/PhysRevLett.118.030401}
  {\bibfield  {journal} {\bibinfo  {journal} {Phys. Rev. Lett.}\ }\textbf
  {\bibinfo {volume} {118}},\ \bibinfo {pages} {030401} (\bibinfo {year}
  {2017})}\BibitemShut {NoStop}%
\bibitem [{\citenamefont {Else}\ \emph {et~al.}(2020)\citenamefont {Else},
  \citenamefont {Ho},\ and\ \citenamefont {Dumitrescu}}]{else2020longLived}%
  \BibitemOpen
  \bibfield  {author} {\bibinfo {author} {\bibfnamefont {D.~V.}\ \bibnamefont
  {Else}}, \bibinfo {author} {\bibfnamefont {W.~W.}\ \bibnamefont {Ho}},\ and\
  \bibinfo {author} {\bibfnamefont {P.~T.}\ \bibnamefont {Dumitrescu}},\
  }\bibfield  {title} {\bibinfo {title} {Long-lived interacting phases of
  matter protected by multiple time-translation symmetries in quasiperiodically
  driven systems},\ }\href {https://doi.org/10.1103/PhysRevX.10.021032}
  {\bibfield  {journal} {\bibinfo  {journal} {Phys. Rev. X}\ }\textbf {\bibinfo
  {volume} {10}},\ \bibinfo {pages} {021032} (\bibinfo {year}
  {2020})}\BibitemShut {NoStop}%
\bibitem [{\citenamefont {Choudhury}(2021)}]{choudhury2021route}%
  \BibitemOpen
  \bibfield  {author} {\bibinfo {author} {\bibfnamefont {S.}~\bibnamefont
  {Choudhury}},\ }\bibfield  {title} {\bibinfo {title} {Route to extend the
  lifetime of a discrete time crystal in a finite spin chain without
  disorder},\ }\href@noop {} {\bibfield  {journal} {\bibinfo  {journal}
  {Atoms}\ }\textbf {\bibinfo {volume} {9}},\ \bibinfo {pages} {25} (\bibinfo
  {year} {2021})}\BibitemShut {NoStop}%
\bibitem [{\citenamefont {Chen}\ \emph {et~al.}(2024)\citenamefont {Chen},
  \citenamefont {Peng}, \citenamefont {Li}, \citenamefont {Chesi},\ and\
  \citenamefont {Wang}}]{Chen2024}%
  \BibitemOpen
  \bibfield  {author} {\bibinfo {author} {\bibfnamefont {D.}~\bibnamefont
  {Chen}}, \bibinfo {author} {\bibfnamefont {Z.}~\bibnamefont {Peng}}, \bibinfo
  {author} {\bibfnamefont {J.}~\bibnamefont {Li}}, \bibinfo {author}
  {\bibfnamefont {S.}~\bibnamefont {Chesi}},\ and\ \bibinfo {author}
  {\bibfnamefont {Y.}~\bibnamefont {Wang}},\ }\bibfield  {title} {\bibinfo
  {title} {Discrete time crystal in an open optomechanical system},\ }\href
  {https://doi.org/10.1103/PhysRevResearch.6.013130} {\bibfield  {journal}
  {\bibinfo  {journal} {Phys. Rev. Res.}\ }\textbf {\bibinfo {volume} {6}},\
  \bibinfo {pages} {013130} (\bibinfo {year} {2024})}\BibitemShut {NoStop}%
\bibitem [{\citenamefont {Schweizer}\ \emph {et~al.}(2019)\citenamefont
  {Schweizer}, \citenamefont {Grusdt}, \citenamefont {Berngruber},
  \citenamefont {Barbiero}, \citenamefont {Demler}, \citenamefont {Goldman},
  \citenamefont {Bloch},\ and\ \citenamefont
  {Aidelsburger}}]{schweizer2019floquet}%
  \BibitemOpen
  \bibfield  {author} {\bibinfo {author} {\bibfnamefont {C.}~\bibnamefont
  {Schweizer}}, \bibinfo {author} {\bibfnamefont {F.}~\bibnamefont {Grusdt}},
  \bibinfo {author} {\bibfnamefont {M.}~\bibnamefont {Berngruber}}, \bibinfo
  {author} {\bibfnamefont {L.}~\bibnamefont {Barbiero}}, \bibinfo {author}
  {\bibfnamefont {E.}~\bibnamefont {Demler}}, \bibinfo {author} {\bibfnamefont
  {N.}~\bibnamefont {Goldman}}, \bibinfo {author} {\bibfnamefont
  {I.}~\bibnamefont {Bloch}},\ and\ \bibinfo {author} {\bibfnamefont
  {M.}~\bibnamefont {Aidelsburger}},\ }\bibfield  {title} {\bibinfo {title}
  {Floquet approach to $\mathbb{Z}_2$ lattice gauge theories with ultracold
  atoms in optical lattices},\ }\href@noop {} {\bibfield  {journal} {\bibinfo
  {journal} {Nature Physics}\ }\textbf {\bibinfo {volume} {15}},\ \bibinfo
  {pages} {1168} (\bibinfo {year} {2019})}\BibitemShut {NoStop}%
\bibitem [{\citenamefont {\ifmmode \check{C}\else
  \v{C}\fi{}ade\ifmmode~\check{z}\else \v{z}\fi{}}\ \emph
  {et~al.}(2019)\citenamefont {\ifmmode \check{C}\else
  \v{C}\fi{}ade\ifmmode~\check{z}\else \v{z}\fi{}}, \citenamefont {Mondaini},\
  and\ \citenamefont {Sacramento}}]{cadez2019}%
  \BibitemOpen
  \bibfield  {author} {\bibinfo {author} {\bibfnamefont {T.}~\bibnamefont
  {\ifmmode \check{C}\else \v{C}\fi{}ade\ifmmode~\check{z}\else \v{z}\fi{}}},
  \bibinfo {author} {\bibfnamefont {R.}~\bibnamefont {Mondaini}},\ and\
  \bibinfo {author} {\bibfnamefont {P.~D.}\ \bibnamefont {Sacramento}},\
  }\bibfield  {title} {\bibinfo {title} {Edge and bulk localization of floquet
  topological superconductors},\ }\href
  {https://doi.org/10.1103/PhysRevB.99.014301} {\bibfield  {journal} {\bibinfo
  {journal} {Phys. Rev. B}\ }\textbf {\bibinfo {volume} {99}},\ \bibinfo
  {pages} {014301} (\bibinfo {year} {2019})}\BibitemShut {NoStop}%
\bibitem [{\citenamefont {Navarrete-Benlloch}\ \emph
  {et~al.}(2021)\citenamefont {Navarrete-Benlloch}, \citenamefont {Garc\'es},
  \citenamefont {Mohseni},\ and\ \citenamefont
  {de~Valc\'arcel}}]{NavarreteBenlloch2021}%
  \BibitemOpen
  \bibfield  {author} {\bibinfo {author} {\bibfnamefont {C.}~\bibnamefont
  {Navarrete-Benlloch}}, \bibinfo {author} {\bibfnamefont {R.}~\bibnamefont
  {Garc\'es}}, \bibinfo {author} {\bibfnamefont {N.}~\bibnamefont {Mohseni}},\
  and\ \bibinfo {author} {\bibfnamefont {G.~J.}\ \bibnamefont
  {de~Valc\'arcel}},\ }\bibfield  {title} {\bibinfo {title} {Floquet theory for
  temporal correlations and spectra in time-periodic open quantum systems:
  Application to squeezed parametric oscillation beyond the rotating-wave
  approximation},\ }\href {https://doi.org/10.1103/PhysRevA.103.023713}
  {\bibfield  {journal} {\bibinfo  {journal} {Phys. Rev. A}\ }\textbf {\bibinfo
  {volume} {103}},\ \bibinfo {pages} {023713} (\bibinfo {year}
  {2021})}\BibitemShut {NoStop}%
\bibitem [{\citenamefont {Yan}\ \emph {et~al.}(2023)\citenamefont {Yan},
  \citenamefont {Lü}, \citenamefont {Luo},\ and\ \citenamefont
  {Zheng}}]{Yan2023}%
  \BibitemOpen
  \bibfield  {author} {\bibinfo {author} {\bibfnamefont {Y.}~\bibnamefont
  {Yan}}, \bibinfo {author} {\bibfnamefont {Z.}~\bibnamefont {Lü}}, \bibinfo
  {author} {\bibfnamefont {J.}~\bibnamefont {Luo}},\ and\ \bibinfo {author}
  {\bibfnamefont {H.}~\bibnamefont {Zheng}},\ }\bibfield  {title} {\bibinfo
  {title} {Controllable sidebands of resonance fluorescence of a two-level
  system driven by bichromatic field},\ }\href
  {https://doi.org/10.1088/1402-4896/acc98b} {\bibfield  {journal} {\bibinfo
  {journal} {Physica Scripta}\ }\textbf {\bibinfo {volume} {98}},\ \bibinfo
  {pages} {055103} (\bibinfo {year} {2023})}\BibitemShut {NoStop}%
\bibitem [{\citenamefont {Kolisnyk}\ \emph {et~al.}(2024)\citenamefont
  {Kolisnyk}, \citenamefont {Quei\ss{}er}, \citenamefont {Schaller},\ and\
  \citenamefont {Sch\"utzhold}}]{Kolisnyk2024}%
  \BibitemOpen
  \bibfield  {author} {\bibinfo {author} {\bibfnamefont {D.}~\bibnamefont
  {Kolisnyk}}, \bibinfo {author} {\bibfnamefont {F.}~\bibnamefont
  {Quei\ss{}er}}, \bibinfo {author} {\bibfnamefont {G.}~\bibnamefont
  {Schaller}},\ and\ \bibinfo {author} {\bibfnamefont {R.}~\bibnamefont
  {Sch\"utzhold}},\ }\bibfield  {title} {\bibinfo {title} {Floquet analysis of
  a superradiant many-qutrit refrigerator},\ }\href
  {https://doi.org/10.1103/PhysRevApplied.21.044050} {\bibfield  {journal}
  {\bibinfo  {journal} {Phys. Rev. Appl.}\ }\textbf {\bibinfo {volume} {21}},\
  \bibinfo {pages} {044050} (\bibinfo {year} {2024})}\BibitemShut {NoStop}%
\bibitem [{\citenamefont {Kohler}(2024)}]{Kohler2024}%
  \BibitemOpen
  \bibfield  {author} {\bibinfo {author} {\bibfnamefont {S.}~\bibnamefont
  {Kohler}},\ }\href@noop {} {\bibinfo {title} {Quantum dissipation at conical
  intersections of quasienergies}} (\bibinfo {year} {2024}),\ \Eprint
  {https://arxiv.org/abs/2405.12093} {arXiv:2405.12093 [cond-mat.mes-hall]}
  \BibitemShut {NoStop}%
\bibitem [{\citenamefont {Zhang}\ \emph {et~al.}(2023)\citenamefont {Zhang},
  \citenamefont {Wang}, \citenamefont {Wang}, \citenamefont {Zhang},
  \citenamefont {Wu}, \citenamefont {Jie},\ and\ \citenamefont
  {Lu}}]{Zhang2023}%
  \BibitemOpen
  \bibfield  {author} {\bibinfo {author} {\bibfnamefont {L.}~\bibnamefont
  {Zhang}}, \bibinfo {author} {\bibfnamefont {Z.}~\bibnamefont {Wang}},
  \bibinfo {author} {\bibfnamefont {Y.}~\bibnamefont {Wang}}, \bibinfo {author}
  {\bibfnamefont {J.}~\bibnamefont {Zhang}}, \bibinfo {author} {\bibfnamefont
  {Z.}~\bibnamefont {Wu}}, \bibinfo {author} {\bibfnamefont {J.}~\bibnamefont
  {Jie}},\ and\ \bibinfo {author} {\bibfnamefont {Y.}~\bibnamefont {Lu}},\
  }\bibfield  {title} {\bibinfo {title} {Quantum synchronization of a single
  trapped-ion qubit},\ }\href
  {https://doi.org/10.1103/PhysRevResearch.5.033209} {\bibfield  {journal}
  {\bibinfo  {journal} {Phys. Rev. Res.}\ }\textbf {\bibinfo {volume} {5}},\
  \bibinfo {pages} {033209} (\bibinfo {year} {2023})}\BibitemShut {NoStop}%
\bibitem [{\citenamefont {Scully}\ and\ \citenamefont
  {Zubairy}(1997)}]{Scully1997}%
  \BibitemOpen
  \bibfield  {author} {\bibinfo {author} {\bibfnamefont {M.~O.}\ \bibnamefont
  {Scully}}\ and\ \bibinfo {author} {\bibfnamefont {M.~S.}\ \bibnamefont
  {Zubairy}},\ }\href@noop {} {\emph {\bibinfo {title} {Quantum optics}}}\
  (\bibinfo  {publisher} {Cambridge Unversity Press},\ \bibinfo {year}
  {1997})\BibitemShut {NoStop}%
\bibitem [{\citenamefont {L.}\ and\ \citenamefont {M.}(2008)}]{Mandel2008}%
  \BibitemOpen
  \bibfield  {author} {\bibinfo {author} {\bibfnamefont {M.}~\bibnamefont
  {L.}}\ and\ \bibinfo {author} {\bibfnamefont {W.}~\bibnamefont {M.}},\
  }\href@noop {} {\emph {\bibinfo {title} {Optical coherence and quantum
  optics}}}\ (\bibinfo  {publisher} {Cambridge University Press, New York},\
  \bibinfo {year} {2008})\BibitemShut {NoStop}%
\bibitem [{\citenamefont {Gardiner}\ and\ \citenamefont
  {Zoller}(2004)}]{Gardiner2004}%
  \BibitemOpen
  \bibfield  {author} {\bibinfo {author} {\bibfnamefont {C.~W.}\ \bibnamefont
  {Gardiner}}\ and\ \bibinfo {author} {\bibfnamefont {P.}~\bibnamefont
  {Zoller}},\ }\href@noop {} {\emph {\bibinfo {title} {Quantum Noise: A
  Handbook of Markovian and Non-Markovian Quantum Stochastic Methods with
  Applications to Quantum Optics}}}\ (\bibinfo  {publisher} {Springer, Berlin,
  Heidelberg},\ \bibinfo {year} {2004})\BibitemShut {NoStop}%
\bibitem [{\citenamefont {Guérin}\ \emph {et~al.}(1997)\citenamefont
  {Guérin}, \citenamefont {Monti}, \citenamefont {Dupont},\ and\ \citenamefont
  {Jauslin}}]{Guerin1997}%
  \BibitemOpen
  \bibfield  {author} {\bibinfo {author} {\bibfnamefont {S.}~\bibnamefont
  {Guérin}}, \bibinfo {author} {\bibfnamefont {F.}~\bibnamefont {Monti}},
  \bibinfo {author} {\bibfnamefont {J.-M.}\ \bibnamefont {Dupont}},\ and\
  \bibinfo {author} {\bibfnamefont {H.~R.}\ \bibnamefont {Jauslin}},\
  }\bibfield  {title} {\bibinfo {title} {{On the relation between
  cavity-dressed states, Floquet states, RWA and semiclassical models}},\
  }\href@noop {} {\bibfield  {journal} {\bibinfo  {journal} {J. Phys. A: Math.
  Gen.}\ }\textbf {\bibinfo {volume} {30}},\ \bibinfo {pages} {7193} (\bibinfo
  {year} {1997})}\BibitemShut {NoStop}%
\bibitem [{\citenamefont {Sambe}(1973)}]{Sambe1973}%
  \BibitemOpen
  \bibfield  {author} {\bibinfo {author} {\bibfnamefont {H.}~\bibnamefont
  {Sambe}},\ }\bibfield  {title} {\bibinfo {title} {Steady states and
  quasienergies of a quantum-mechanical system in an oscillating field},\
  }\href {https://doi.org/10.1103/PhysRevA.7.2203} {\bibfield  {journal}
  {\bibinfo  {journal} {Phys. Rev. A}\ }\textbf {\bibinfo {volume} {7}},\
  \bibinfo {pages} {2203} (\bibinfo {year} {1973})}\BibitemShut {NoStop}%
\bibitem [{\citenamefont {Long}\ \emph {et~al.}(2021)\citenamefont {Long},
  \citenamefont {Crowley},\ and\ \citenamefont {Chandran}}]{Long2021}%
  \BibitemOpen
  \bibfield  {author} {\bibinfo {author} {\bibfnamefont {D.~M.}\ \bibnamefont
  {Long}}, \bibinfo {author} {\bibfnamefont {P.~J.~D.}\ \bibnamefont
  {Crowley}},\ and\ \bibinfo {author} {\bibfnamefont {A.}~\bibnamefont
  {Chandran}},\ }\bibfield  {title} {\bibinfo {title} {Nonadiabatic topological
  energy pumps with quasiperiodic driving},\ }\href
  {https://doi.org/10.1103/PhysRevLett.126.106805} {\bibfield  {journal}
  {\bibinfo  {journal} {Phys. Rev. Lett.}\ }\textbf {\bibinfo {volume} {126}},\
  \bibinfo {pages} {106805} (\bibinfo {year} {2021})}\BibitemShut {NoStop}%
\bibitem [{\citenamefont {Crowley}\ \emph {et~al.}(2019)\citenamefont
  {Crowley}, \citenamefont {Martin},\ and\ \citenamefont
  {Chandran}}]{Crowley2019}%
  \BibitemOpen
  \bibfield  {author} {\bibinfo {author} {\bibfnamefont {P.~J.~D.}\
  \bibnamefont {Crowley}}, \bibinfo {author} {\bibfnamefont {I.}~\bibnamefont
  {Martin}},\ and\ \bibinfo {author} {\bibfnamefont {A.}~\bibnamefont
  {Chandran}},\ }\bibfield  {title} {\bibinfo {title} {Topological
  classification of quasiperiodically driven quantum systems},\ }\href
  {https://doi.org/10.1103/PhysRevB.99.064306} {\bibfield  {journal} {\bibinfo
  {journal} {Phys. Rev. B}\ }\textbf {\bibinfo {volume} {99}},\ \bibinfo
  {pages} {064306} (\bibinfo {year} {2019})}\BibitemShut {NoStop}%
\bibitem [{\citenamefont {Crowley}\ \emph {et~al.}(2020)\citenamefont
  {Crowley}, \citenamefont {Martin},\ and\ \citenamefont
  {Chandran}}]{Crowley2020}%
  \BibitemOpen
  \bibfield  {author} {\bibinfo {author} {\bibfnamefont {P.~J.~D.}\
  \bibnamefont {Crowley}}, \bibinfo {author} {\bibfnamefont {I.}~\bibnamefont
  {Martin}},\ and\ \bibinfo {author} {\bibfnamefont {A.}~\bibnamefont
  {Chandran}},\ }\bibfield  {title} {\bibinfo {title} {Half-integer quantized
  topological response in quasiperiodically driven quantum systems},\ }\href
  {https://doi.org/10.1103/PhysRevLett.125.100601} {\bibfield  {journal}
  {\bibinfo  {journal} {Phys. Rev. Lett.}\ }\textbf {\bibinfo {volume} {125}},\
  \bibinfo {pages} {100601} (\bibinfo {year} {2020})}\BibitemShut {NoStop}%
\bibitem [{\citenamefont {Seifert}(2012)}]{Seifert2012}%
  \BibitemOpen
  \bibfield  {author} {\bibinfo {author} {\bibfnamefont {U.}~\bibnamefont
  {Seifert}},\ }\bibfield  {title} {\bibinfo {title} {Stochastic
  thermodynamics, fluctuation theorems and molecular machines},\ }\href@noop {}
  {\bibfield  {journal} {\bibinfo  {journal} {Reports on Progress in Physics}\
  }\textbf {\bibinfo {volume} {75}},\ \bibinfo {pages} {126001} (\bibinfo
  {year} {2012})}\BibitemShut {NoStop}%
\bibitem [{\citenamefont {Engelhardt}\ \emph
  {et~al.}(2024{\natexlab{a}})\citenamefont {Engelhardt}, \citenamefont
  {Choudhury},\ and\ \citenamefont {Liu}}]{Engelhardt2024}%
  \BibitemOpen
  \bibfield  {author} {\bibinfo {author} {\bibfnamefont {G.}~\bibnamefont
  {Engelhardt}}, \bibinfo {author} {\bibfnamefont {S.}~\bibnamefont
  {Choudhury}},\ and\ \bibinfo {author} {\bibfnamefont {W.~V.}\ \bibnamefont
  {Liu}},\ }\bibfield  {title} {\bibinfo {title} {Unified light-matter floquet
  theory and its application to quantum communication},\ }\href
  {https://doi.org/10.1103/PhysRevResearch.6.013116} {\bibfield  {journal}
  {\bibinfo  {journal} {Phys. Rev. Res.}\ }\textbf {\bibinfo {volume} {6}},\
  \bibinfo {pages} {013116} (\bibinfo {year} {2024}{\natexlab{a}})}\BibitemShut
  {NoStop}%
\bibitem [{\citenamefont {Levitov}\ \emph {et~al.}(1996)\citenamefont
  {Levitov}, \citenamefont {Lee},\ and\ \citenamefont
  {Lesovik}}]{levitov1996electron}%
  \BibitemOpen
  \bibfield  {author} {\bibinfo {author} {\bibfnamefont {L.~S.}\ \bibnamefont
  {Levitov}}, \bibinfo {author} {\bibfnamefont {H.}~\bibnamefont {Lee}},\ and\
  \bibinfo {author} {\bibfnamefont {G.~B.}\ \bibnamefont {Lesovik}},\
  }\bibfield  {title} {\bibinfo {title} {Electron counting statistics and
  coherent states of electric current},\ }\href@noop {} {\bibfield  {journal}
  {\bibinfo  {journal} {Journal of Mathematical Physics}\ }\textbf {\bibinfo
  {volume} {37}},\ \bibinfo {pages} {4845} (\bibinfo {year}
  {1996})}\BibitemShut {NoStop}%
\bibitem [{\citenamefont {Brandes}\ \emph {et~al.}(2004)\citenamefont
  {Brandes}, \citenamefont {Aguado},\ and\ \citenamefont
  {Platero}}]{Brandes2004}%
  \BibitemOpen
  \bibfield  {author} {\bibinfo {author} {\bibfnamefont {T.}~\bibnamefont
  {Brandes}}, \bibinfo {author} {\bibfnamefont {R.}~\bibnamefont {Aguado}},\
  and\ \bibinfo {author} {\bibfnamefont {G.}~\bibnamefont {Platero}},\
  }\bibfield  {title} {\bibinfo {title} {Charge transport through open driven
  two-level systems with dissipation},\ }\href
  {https://doi.org/10.1103/PhysRevB.69.205326} {\bibfield  {journal} {\bibinfo
  {journal} {Phys. Rev. B}\ }\textbf {\bibinfo {volume} {69}},\ \bibinfo
  {pages} {205326} (\bibinfo {year} {2004})}\BibitemShut {NoStop}%
\bibitem [{\citenamefont {Sch\"onhammer}(2007)}]{Schoenhammer2007}%
  \BibitemOpen
  \bibfield  {author} {\bibinfo {author} {\bibfnamefont {K.}~\bibnamefont
  {Sch\"onhammer}},\ }\bibfield  {title} {\bibinfo {title} {{Full counting
  statistics for noninteracting fermions: Exact results and the Levitov-Lesovik
  formula}},\ }\href {https://doi.org/10.1103/PhysRevB.75.205329} {\bibfield
  {journal} {\bibinfo  {journal} {Phys. Rev. B}\ }\textbf {\bibinfo {volume}
  {75}},\ \bibinfo {pages} {205329} (\bibinfo {year} {2007})}\BibitemShut
  {NoStop}%
\bibitem [{\citenamefont {Brandes}(2008)}]{Brandes2008}%
  \BibitemOpen
  \bibfield  {author} {\bibinfo {author} {\bibfnamefont {T.}~\bibnamefont
  {Brandes}},\ }\bibfield  {title} {\bibinfo {title} {Waiting times and noise
  in single particle transport},\ }\href@noop {} {\bibfield  {journal}
  {\bibinfo  {journal} {Ann. Phys. (Berlin)}\ }\textbf {\bibinfo {volume}
  {17}},\ \bibinfo {pages} {477} (\bibinfo {year} {2008})}\BibitemShut
  {NoStop}%
\bibitem [{\citenamefont {Bagrets}\ and\ \citenamefont
  {Nazarov}(2003)}]{Bagrets2003}%
  \BibitemOpen
  \bibfield  {author} {\bibinfo {author} {\bibfnamefont {D.~A.}\ \bibnamefont
  {Bagrets}}\ and\ \bibinfo {author} {\bibfnamefont {Y.~V.}\ \bibnamefont
  {Nazarov}},\ }\bibfield  {title} {\bibinfo {title} {Full counting statistics
  of charge transfer in coulomb blockade systems},\ }\href
  {https://doi.org/10.1103/PhysRevB.67.085316} {\bibfield  {journal} {\bibinfo
  {journal} {Phys. Rev. B}\ }\textbf {\bibinfo {volume} {67}},\ \bibinfo
  {pages} {085316} (\bibinfo {year} {2003})}\BibitemShut {NoStop}%
\bibitem [{\citenamefont {S\'anchez}\ \emph {et~al.}(2007)\citenamefont
  {S\'anchez}, \citenamefont {Platero},\ and\ \citenamefont
  {Brandes}}]{Sanchez2007}%
  \BibitemOpen
  \bibfield  {author} {\bibinfo {author} {\bibfnamefont {R.}~\bibnamefont
  {S\'anchez}}, \bibinfo {author} {\bibfnamefont {G.}~\bibnamefont {Platero}},\
  and\ \bibinfo {author} {\bibfnamefont {T.}~\bibnamefont {Brandes}},\
  }\bibfield  {title} {\bibinfo {title} {Resonance fluorescence in transport
  through quantum dots: {N}oise properties},\ }\href
  {https://doi.org/10.1103/PhysRevLett.98.146805} {\bibfield  {journal}
  {\bibinfo  {journal} {Phys. Rev. Lett.}\ }\textbf {\bibinfo {volume} {98}},\
  \bibinfo {pages} {146805} (\bibinfo {year} {2007})}\BibitemShut {NoStop}%
\bibitem [{\citenamefont {S\'anchez}\ \emph {et~al.}(2008)\citenamefont
  {S\'anchez}, \citenamefont {Platero},\ and\ \citenamefont
  {Brandes}}]{Sanchez2008}%
  \BibitemOpen
  \bibfield  {author} {\bibinfo {author} {\bibfnamefont {R.}~\bibnamefont
  {S\'anchez}}, \bibinfo {author} {\bibfnamefont {G.}~\bibnamefont {Platero}},\
  and\ \bibinfo {author} {\bibfnamefont {T.}~\bibnamefont {Brandes}},\
  }\bibfield  {title} {\bibinfo {title} {Resonance fluorescence in driven
  quantum dots: {E}lectron and photon correlations},\ }\href
  {https://doi.org/10.1103/PhysRevB.78.125308} {\bibfield  {journal} {\bibinfo
  {journal} {Phys. Rev. B}\ }\textbf {\bibinfo {volume} {78}},\ \bibinfo
  {pages} {125308} (\bibinfo {year} {2008})}\BibitemShut {NoStop}%
\bibitem [{\citenamefont {Ding}\ \emph {et~al.}(2024)\citenamefont {Ding},
  \citenamefont {Yan}, \citenamefont {Hu}, \citenamefont {Engelhardt},\ and\
  \citenamefont {Luo}}]{Ding2024}%
  \BibitemOpen
  \bibfield  {author} {\bibinfo {author} {\bibfnamefont {Y.}~\bibnamefont
  {Ding}}, \bibinfo {author} {\bibfnamefont {Y.}~\bibnamefont {Yan}}, \bibinfo
  {author} {\bibfnamefont {J.}~\bibnamefont {Hu}}, \bibinfo {author}
  {\bibfnamefont {G.}~\bibnamefont {Engelhardt}},\ and\ \bibinfo {author}
  {\bibfnamefont {J.}~\bibnamefont {Luo}},\ }\bibfield  {title} {\bibinfo
  {title} {Anomalous conditional counting statistics in an
  electron-spin-resonance quantum dot measured by a quantum point contact},\
  }\href {https://doi.org/10.1103/PhysRevB.109.115136} {\bibfield  {journal}
  {\bibinfo  {journal} {Phys. Rev. B}\ }\textbf {\bibinfo {volume} {109}},\
  \bibinfo {pages} {115136} (\bibinfo {year} {2024})}\BibitemShut {NoStop}%
\bibitem [{\citenamefont {Xu}\ \emph {et~al.}(2023)\citenamefont {Xu},
  \citenamefont {Wang}, \citenamefont {Wu}, \citenamefont {Yan}, \citenamefont
  {Hu}, \citenamefont {Engelhardt},\ and\ \citenamefont {Luo}}]{Xu2023}%
  \BibitemOpen
  \bibfield  {author} {\bibinfo {author} {\bibfnamefont {J.}~\bibnamefont
  {Xu}}, \bibinfo {author} {\bibfnamefont {S.}~\bibnamefont {Wang}}, \bibinfo
  {author} {\bibfnamefont {J.}~\bibnamefont {Wu}}, \bibinfo {author}
  {\bibfnamefont {Y.}~\bibnamefont {Yan}}, \bibinfo {author} {\bibfnamefont
  {J.}~\bibnamefont {Hu}}, \bibinfo {author} {\bibfnamefont {G.}~\bibnamefont
  {Engelhardt}},\ and\ \bibinfo {author} {\bibfnamefont {J.}~\bibnamefont
  {Luo}},\ }\bibfield  {title} {\bibinfo {title} {Noise suppression of
  transport through double quantum dots by feedback control},\ }\href
  {https://doi.org/10.1103/PhysRevB.107.125113} {\bibfield  {journal} {\bibinfo
   {journal} {Phys. Rev. B}\ }\textbf {\bibinfo {volume} {107}},\ \bibinfo
  {pages} {125113} (\bibinfo {year} {2023})}\BibitemShut {NoStop}%
\bibitem [{\citenamefont {Restrepo}\ \emph {et~al.}(2018)\citenamefont
  {Restrepo}, \citenamefont {Cerrillo}, \citenamefont {Strasberg},\ and\
  \citenamefont {Schaller}}]{Restrepo2018}%
  \BibitemOpen
  \bibfield  {author} {\bibinfo {author} {\bibfnamefont {S.}~\bibnamefont
  {Restrepo}}, \bibinfo {author} {\bibfnamefont {J.}~\bibnamefont {Cerrillo}},
  \bibinfo {author} {\bibfnamefont {P.}~\bibnamefont {Strasberg}},\ and\
  \bibinfo {author} {\bibfnamefont {G.}~\bibnamefont {Schaller}},\ }\bibfield
  {title} {\bibinfo {title} {From quantum heat engines to laser cooling:
  {Floque}t theory beyond the {Born}{\textendash}{Markov} approximation},\
  }\href {https://doi.org/10.1088/1367-2630/aac583} {\bibfield  {journal}
  {\bibinfo  {journal} {New Journal of Physics}\ }\textbf {\bibinfo {volume}
  {20}},\ \bibinfo {pages} {053063} (\bibinfo {year} {2018})}\BibitemShut
  {NoStop}%
\bibitem [{\citenamefont {Schaller}\ \emph {et~al.}(2018)\citenamefont
  {Schaller}, \citenamefont {Cerrillo}, \citenamefont {Engelhardt},\ and\
  \citenamefont {Strasberg}}]{Schaller2018}%
  \BibitemOpen
  \bibfield  {author} {\bibinfo {author} {\bibfnamefont {G.}~\bibnamefont
  {Schaller}}, \bibinfo {author} {\bibfnamefont {J.}~\bibnamefont {Cerrillo}},
  \bibinfo {author} {\bibfnamefont {G.}~\bibnamefont {Engelhardt}},\ and\
  \bibinfo {author} {\bibfnamefont {P.}~\bibnamefont {Strasberg}},\ }\bibfield
  {title} {\bibinfo {title} {{Electronic Maxwell demon in the coherent
  strong-coupling regime}},\ }\href
  {https://doi.org/10.1103/PhysRevB.97.195104} {\bibfield  {journal} {\bibinfo
  {journal} {Phys. Rev. B}\ }\textbf {\bibinfo {volume} {97}},\ \bibinfo
  {pages} {195104} (\bibinfo {year} {2018})}\BibitemShut {NoStop}%
\bibitem [{\citenamefont {Cao}\ \emph {et~al.}(2023)\citenamefont {Cao},
  \citenamefont {Wang},\ and\ \citenamefont {He}}]{Cao2023}%
  \BibitemOpen
  \bibfield  {author} {\bibinfo {author} {\bibfnamefont {X.}~\bibnamefont
  {Cao}}, \bibinfo {author} {\bibfnamefont {C.}~\bibnamefont {Wang}},\ and\
  \bibinfo {author} {\bibfnamefont {D.}~\bibnamefont {He}},\ }\bibfield
  {title} {\bibinfo {title} {Driving induced coherent quantum energy
  transport},\ }\href {https://doi.org/10.1103/PhysRevB.108.245401} {\bibfield
  {journal} {\bibinfo  {journal} {Phys. Rev. B}\ }\textbf {\bibinfo {volume}
  {108}},\ \bibinfo {pages} {245401} (\bibinfo {year} {2023})}\BibitemShut
  {NoStop}%
\bibitem [{\citenamefont {Wang}\ and\ \citenamefont {Ren}(2024)}]{Wang2024}%
  \BibitemOpen
  \bibfield  {author} {\bibinfo {author} {\bibfnamefont {Z.}~\bibnamefont
  {Wang}}\ and\ \bibinfo {author} {\bibfnamefont {J.}~\bibnamefont {Ren}},\
  }\bibfield  {title} {\bibinfo {title} {Thermodynamic geometry of
  nonequilibrium fluctuations in cyclically driven transport},\ }\href
  {https://doi.org/10.1103/PhysRevLett.132.207101} {\bibfield  {journal}
  {\bibinfo  {journal} {Phys. Rev. Lett.}\ }\textbf {\bibinfo {volume} {132}},\
  \bibinfo {pages} {207101} (\bibinfo {year} {2024})}\BibitemShut {NoStop}%
\bibitem [{\citenamefont {Engelhardt}\ \emph {et~al.}(2023)\citenamefont
  {Engelhardt}, \citenamefont {Luo}, \citenamefont {Bastidas},\ and\
  \citenamefont {Platero}}]{Engelhardt2024c}%
  \BibitemOpen
  \bibfield  {author} {\bibinfo {author} {\bibfnamefont {G.}~\bibnamefont
  {Engelhardt}}, \bibinfo {author} {\bibfnamefont {J.}~\bibnamefont {Luo}},
  \bibinfo {author} {\bibfnamefont {V.~M.}\ \bibnamefont {Bastidas}},\ and\
  \bibinfo {author} {\bibfnamefont {G.}~\bibnamefont {Platero}},\ }\bibfield
  {title} {\bibinfo {title} {{Photon-resolved Floquet theory II: open quantum
  systems}},\ }\href@noop {} {\bibfield  {journal} {\bibinfo  {journal}
  {arXiv:2311.01509}\ } (\bibinfo {year} {2023})}\BibitemShut {NoStop}%
\bibitem [{\citenamefont {Yuen-Zhou}\ and\ \citenamefont
  {Menon}(2019)}]{YuenZhou2019}%
  \BibitemOpen
  \bibfield  {author} {\bibinfo {author} {\bibfnamefont {J.}~\bibnamefont
  {Yuen-Zhou}}\ and\ \bibinfo {author} {\bibfnamefont {V.~M.}\ \bibnamefont
  {Menon}},\ }\bibfield  {title} {\bibinfo {title} {Polariton chemistry:
  Thinking inside the (photon) box},\ }\href@noop {} {\bibfield  {journal}
  {\bibinfo  {journal} {Proceedings of the National Academy of Sciences}\
  }\textbf {\bibinfo {volume} {116}},\ \bibinfo {pages} {5214} (\bibinfo {year}
  {2019})}\BibitemShut {NoStop}%
\bibitem [{\citenamefont {Herrera}\ and\ \citenamefont
  {Owrutsky}(2020)}]{Herrera2020}%
  \BibitemOpen
  \bibfield  {author} {\bibinfo {author} {\bibfnamefont {F.}~\bibnamefont
  {Herrera}}\ and\ \bibinfo {author} {\bibfnamefont {J.}~\bibnamefont
  {Owrutsky}},\ }\bibfield  {title} {\bibinfo {title} {Molecular polaritons for
  controlling chemistry with quantum optics},\ }\href
  {https://doi.org/10.1063/1.5136320} {\bibfield  {journal} {\bibinfo
  {journal} {The Journal of Chemical Physics}\ }\textbf {\bibinfo {volume}
  {152}},\ \bibinfo {pages} {100902} (\bibinfo {year} {2020})}\BibitemShut
  {NoStop}%
\bibitem [{\citenamefont {Ribeiro}\ \emph {et~al.}(2018)\citenamefont
  {Ribeiro}, \citenamefont {Martínez-Martínez}, \citenamefont {Du},
  \citenamefont {Campos-Gonzalez-Angulo},\ and\ \citenamefont
  {Yuen-Zhou}}]{Ribeiro2018}%
  \BibitemOpen
  \bibfield  {author} {\bibinfo {author} {\bibfnamefont {R.~F.}\ \bibnamefont
  {Ribeiro}}, \bibinfo {author} {\bibfnamefont {L.~A.}\ \bibnamefont
  {Martínez-Martínez}}, \bibinfo {author} {\bibfnamefont {M.}~\bibnamefont
  {Du}}, \bibinfo {author} {\bibfnamefont {J.}~\bibnamefont
  {Campos-Gonzalez-Angulo}},\ and\ \bibinfo {author} {\bibfnamefont
  {J.}~\bibnamefont {Yuen-Zhou}},\ }\bibfield  {title} {\bibinfo {title}
  {Polariton chemistry: controlling molecular dynamics with optical cavities},\
  }\href {https://doi.org/10.1039/C8SC01043A} {\bibfield  {journal} {\bibinfo
  {journal} {Chem. Sci.}\ }\textbf {\bibinfo {volume} {9}},\ \bibinfo {pages}
  {6325} (\bibinfo {year} {2018})}\BibitemShut {NoStop}%
\bibitem [{\citenamefont {Ohta}\ \emph {et~al.}(2024)\citenamefont {Ohta},
  \citenamefont {Lelu}, \citenamefont {Xu}, \citenamefont {Inaba},
  \citenamefont {Hitachi}, \citenamefont {Taniyasu}, \citenamefont {Sanada},
  \citenamefont {Ishizawa}, \citenamefont {Tawara}, \citenamefont {Oguri},
  \citenamefont {Yamaguchi},\ and\ \citenamefont {Okamoto}}]{Ohta2024}%
  \BibitemOpen
  \bibfield  {author} {\bibinfo {author} {\bibfnamefont {R.}~\bibnamefont
  {Ohta}}, \bibinfo {author} {\bibfnamefont {G.}~\bibnamefont {Lelu}}, \bibinfo
  {author} {\bibfnamefont {X.}~\bibnamefont {Xu}}, \bibinfo {author}
  {\bibfnamefont {T.}~\bibnamefont {Inaba}}, \bibinfo {author} {\bibfnamefont
  {K.}~\bibnamefont {Hitachi}}, \bibinfo {author} {\bibfnamefont
  {Y.}~\bibnamefont {Taniyasu}}, \bibinfo {author} {\bibfnamefont
  {H.}~\bibnamefont {Sanada}}, \bibinfo {author} {\bibfnamefont
  {A.}~\bibnamefont {Ishizawa}}, \bibinfo {author} {\bibfnamefont
  {T.}~\bibnamefont {Tawara}}, \bibinfo {author} {\bibfnamefont
  {K.}~\bibnamefont {Oguri}}, \bibinfo {author} {\bibfnamefont
  {H.}~\bibnamefont {Yamaguchi}},\ and\ \bibinfo {author} {\bibfnamefont
  {H.}~\bibnamefont {Okamoto}},\ }\bibfield  {title} {\bibinfo {title}
  {Observation of acoustically induced dressed states of rare-earth ions},\
  }\href {https://doi.org/10.1103/PhysRevLett.132.036904} {\bibfield  {journal}
  {\bibinfo  {journal} {Phys. Rev. Lett.}\ }\textbf {\bibinfo {volume} {132}},\
  \bibinfo {pages} {036904} (\bibinfo {year} {2024})}\BibitemShut {NoStop}%
\bibitem [{\citenamefont {Panahiyan}\ \emph {et~al.}(2023)\citenamefont
  {Panahiyan}, \citenamefont {Mu\~noz}, \citenamefont {Chekhova},\ and\
  \citenamefont {Schlawin}}]{Panahiyan2023}%
  \BibitemOpen
  \bibfield  {author} {\bibinfo {author} {\bibfnamefont {S.}~\bibnamefont
  {Panahiyan}}, \bibinfo {author} {\bibfnamefont {C.~S.}\ \bibnamefont
  {Mu\~noz}}, \bibinfo {author} {\bibfnamefont {M.~V.}\ \bibnamefont
  {Chekhova}},\ and\ \bibinfo {author} {\bibfnamefont {F.}~\bibnamefont
  {Schlawin}},\ }\bibfield  {title} {\bibinfo {title} {Nonlinear interferometry
  for quantum-enhanced measurements of multiphoton absorption},\ }\href
  {https://doi.org/10.1103/PhysRevLett.130.203604} {\bibfield  {journal}
  {\bibinfo  {journal} {Phys. Rev. Lett.}\ }\textbf {\bibinfo {volume} {130}},\
  \bibinfo {pages} {203604} (\bibinfo {year} {2023})}\BibitemShut {NoStop}%
\bibitem [{\citenamefont {Dorfman}\ \emph {et~al.}(2021)\citenamefont
  {Dorfman}, \citenamefont {Asban}, \citenamefont {Gu},\ and\ \citenamefont
  {Mukamel}}]{Dorfman2021}%
  \BibitemOpen
  \bibfield  {author} {\bibinfo {author} {\bibfnamefont {K.~E.}\ \bibnamefont
  {Dorfman}}, \bibinfo {author} {\bibfnamefont {S.}~\bibnamefont {Asban}},
  \bibinfo {author} {\bibfnamefont {B.}~\bibnamefont {Gu}},\ and\ \bibinfo
  {author} {\bibfnamefont {S.}~\bibnamefont {Mukamel}},\ }\bibfield  {title}
  {\bibinfo {title} {Hong-ou-mandel interferometry and spectroscopy using
  entangled photons},\ }\href {https://doi.org/10.1038/s42005-021-00542-2}
  {\bibfield  {journal} {\bibinfo  {journal} {Communications Physics}\ }\textbf
  {\bibinfo {volume} {4}},\ \bibinfo {pages} {49} (\bibinfo {year}
  {2021})}\BibitemShut {NoStop}%
\bibitem [{\citenamefont {Zhang}\ \emph {et~al.}(2022)\citenamefont {Zhang},
  \citenamefont {Peng}, \citenamefont {Nie}, \citenamefont {Agarwal},\ and\
  \citenamefont {Scully}}]{Zhang2022}%
  \BibitemOpen
  \bibfield  {author} {\bibinfo {author} {\bibfnamefont {Z.}~\bibnamefont
  {Zhang}}, \bibinfo {author} {\bibfnamefont {T.}~\bibnamefont {Peng}},
  \bibinfo {author} {\bibfnamefont {X.}~\bibnamefont {Nie}}, \bibinfo {author}
  {\bibfnamefont {G.~S.}\ \bibnamefont {Agarwal}},\ and\ \bibinfo {author}
  {\bibfnamefont {M.~O.}\ \bibnamefont {Scully}},\ }\bibfield  {title}
  {\bibinfo {title} {Entangled photons enabled time-frequency-resolved coherent
  {Raman} spectroscopy and applications to electronic coherences at femtosecond
  scale},\ }\bibfield  {journal} {\bibinfo  {journal} {Light: Science and
  Applications}\ }\textbf {\bibinfo {volume} {11}},\ \href
  {https://doi.org/10.1038/s41377-022-00953-y} {10.1038/s41377-022-00953-y}
  (\bibinfo {year} {2022})\BibitemShut {NoStop}%
\bibitem [{\citenamefont {Gao}\ \emph {et~al.}(2021)\citenamefont {Gao},
  \citenamefont {Genevet}, \citenamefont {Li},\ and\ \citenamefont
  {Dorfman}}]{Gao2021}%
  \BibitemOpen
  \bibfield  {author} {\bibinfo {author} {\bibfnamefont {Z.}~\bibnamefont
  {Gao}}, \bibinfo {author} {\bibfnamefont {P.}~\bibnamefont {Genevet}},
  \bibinfo {author} {\bibfnamefont {G.}~\bibnamefont {Li}},\ and\ \bibinfo
  {author} {\bibfnamefont {K.~E.}\ \bibnamefont {Dorfman}},\ }\bibfield
  {title} {\bibinfo {title} {Reconstruction of multidimensional nonlinear
  polarization response of pancharatnam-berry metasurfaces},\ }\href
  {https://doi.org/10.1103/PhysRevB.104.054303} {\bibfield  {journal} {\bibinfo
   {journal} {Phys. Rev. B}\ }\textbf {\bibinfo {volume} {104}},\ \bibinfo
  {pages} {054303} (\bibinfo {year} {2021})}\BibitemShut {NoStop}%
\bibitem [{\citenamefont {Blais}\ \emph {et~al.}(2021)\citenamefont {Blais},
  \citenamefont {Grimsmo}, \citenamefont {Girvin},\ and\ \citenamefont
  {Wallraff}}]{Blais2021}%
  \BibitemOpen
  \bibfield  {author} {\bibinfo {author} {\bibfnamefont {A.}~\bibnamefont
  {Blais}}, \bibinfo {author} {\bibfnamefont {A.~L.}\ \bibnamefont {Grimsmo}},
  \bibinfo {author} {\bibfnamefont {S.~M.}\ \bibnamefont {Girvin}},\ and\
  \bibinfo {author} {\bibfnamefont {A.}~\bibnamefont {Wallraff}},\ }\bibfield
  {title} {\bibinfo {title} {Circuit quantum electrodynamics},\ }\href
  {https://doi.org/10.1103/RevModPhys.93.025005} {\bibfield  {journal}
  {\bibinfo  {journal} {Rev. Mod. Phys.}\ }\textbf {\bibinfo {volume} {93}},\
  \bibinfo {pages} {025005} (\bibinfo {year} {2021})}\BibitemShut {NoStop}%
\bibitem [{\citenamefont {Zhang}\ \emph {et~al.}(2019)\citenamefont {Zhang},
  \citenamefont {Zhou}, \citenamefont {Feng},\ and\ \citenamefont
  {Li}}]{Zhang2019a}%
  \BibitemOpen
  \bibfield  {author} {\bibinfo {author} {\bibfnamefont {C.}~\bibnamefont
  {Zhang}}, \bibinfo {author} {\bibfnamefont {K.}~\bibnamefont {Zhou}},
  \bibinfo {author} {\bibfnamefont {W.}~\bibnamefont {Feng}},\ and\ \bibinfo
  {author} {\bibfnamefont {X.-Q.}\ \bibnamefont {Li}},\ }\bibfield  {title}
  {\bibinfo {title} {Estimation of parameters in circuit qed by continuous
  quantum measurement},\ }\href {https://doi.org/10.1103/PhysRevA.99.022114}
  {\bibfield  {journal} {\bibinfo  {journal} {Phys. Rev. A}\ }\textbf {\bibinfo
  {volume} {99}},\ \bibinfo {pages} {022114} (\bibinfo {year}
  {2019})}\BibitemShut {NoStop}%
\bibitem [{\citenamefont {Bloch}\ \emph {et~al.}(2022)\citenamefont {Bloch},
  \citenamefont {Ronen}, \citenamefont {Shaham}, \citenamefont {Katz},
  \citenamefont {Volansky},\ and\ \citenamefont {Katz}}]{Bloch2022}%
  \BibitemOpen
  \bibfield  {author} {\bibinfo {author} {\bibfnamefont {I.~M.}\ \bibnamefont
  {Bloch}}, \bibinfo {author} {\bibfnamefont {G.}~\bibnamefont {Ronen}},
  \bibinfo {author} {\bibfnamefont {R.}~\bibnamefont {Shaham}}, \bibinfo
  {author} {\bibfnamefont {O.}~\bibnamefont {Katz}}, \bibinfo {author}
  {\bibfnamefont {T.}~\bibnamefont {Volansky}},\ and\ \bibinfo {author}
  {\bibfnamefont {O.}~\bibnamefont {Katz}},\ }\bibfield  {title} {\bibinfo
  {title} {New constraints on axion-like dark matter using a {Floquet} quantum
  detector},\ }\href {https://doi.org/10.1126/sciadv.abl8919} {\bibfield
  {journal} {\bibinfo  {journal} {Science Advances}\ }\textbf {\bibinfo
  {volume} {8}},\ \bibinfo {pages} {eabl8919} (\bibinfo {year}
  {2022})}\BibitemShut {NoStop}%
\bibitem [{\citenamefont {Engelhardt}\ \emph
  {et~al.}(2024{\natexlab{b}})\citenamefont {Engelhardt}, \citenamefont
  {Bhoonah},\ and\ \citenamefont {Liu}}]{Engelhardt2024a}%
  \BibitemOpen
  \bibfield  {author} {\bibinfo {author} {\bibfnamefont {G.}~\bibnamefont
  {Engelhardt}}, \bibinfo {author} {\bibfnamefont {A.}~\bibnamefont
  {Bhoonah}},\ and\ \bibinfo {author} {\bibfnamefont {W.~V.}\ \bibnamefont
  {Liu}},\ }\bibfield  {title} {\bibinfo {title} {Detecting axion dark matter
  with rydberg atoms via induced electric dipole transitions},\ }\href
  {https://doi.org/10.1103/PhysRevResearch.6.023017} {\bibfield  {journal}
  {\bibinfo  {journal} {Phys. Rev. Res.}\ }\textbf {\bibinfo {volume} {6}},\
  \bibinfo {pages} {023017} (\bibinfo {year} {2024}{\natexlab{b}})}\BibitemShut
  {NoStop}%
\bibitem [{\citenamefont {Vigneau}\ \emph {et~al.}(2023)\citenamefont
  {Vigneau}, \citenamefont {Fedele}, \citenamefont {Chatterjee}, \citenamefont
  {Reilly}, \citenamefont {Kuemmeth}, \citenamefont {Gonzalez-Zalba},
  \citenamefont {Laird},\ and\ \citenamefont {Ares}}]{Vigneau2023}%
  \BibitemOpen
  \bibfield  {author} {\bibinfo {author} {\bibfnamefont {F.}~\bibnamefont
  {Vigneau}}, \bibinfo {author} {\bibfnamefont {F.}~\bibnamefont {Fedele}},
  \bibinfo {author} {\bibfnamefont {A.}~\bibnamefont {Chatterjee}}, \bibinfo
  {author} {\bibfnamefont {D.}~\bibnamefont {Reilly}}, \bibinfo {author}
  {\bibfnamefont {F.}~\bibnamefont {Kuemmeth}}, \bibinfo {author}
  {\bibfnamefont {M.~F.}\ \bibnamefont {Gonzalez-Zalba}}, \bibinfo {author}
  {\bibfnamefont {E.}~\bibnamefont {Laird}},\ and\ \bibinfo {author}
  {\bibfnamefont {N.}~\bibnamefont {Ares}},\ }\bibfield  {title} {\bibinfo
  {title} {{Probing quantum devices with radio-frequency reflectometry}},\
  }\href {https://doi.org/10.1063/5.0088229} {\bibfield  {journal} {\bibinfo
  {journal} {Applied Physics Reviews}\ }\textbf {\bibinfo {volume} {10}},\
  \bibinfo {pages} {021305} (\bibinfo {year} {2023})}\BibitemShut {NoStop}%
\bibitem [{\citenamefont {Lecocq}\ \emph {et~al.}(2020)\citenamefont {Lecocq},
  \citenamefont {Ranzani}, \citenamefont {Peterson}, \citenamefont {Cicak},
  \citenamefont {Metelmann}, \citenamefont {Kotler}, \citenamefont {Simmonds},
  \citenamefont {Teufel},\ and\ \citenamefont {Aumentado}}]{Lecocq2020}%
  \BibitemOpen
  \bibfield  {author} {\bibinfo {author} {\bibfnamefont {F.}~\bibnamefont
  {Lecocq}}, \bibinfo {author} {\bibfnamefont {L.}~\bibnamefont {Ranzani}},
  \bibinfo {author} {\bibfnamefont {G.}~\bibnamefont {Peterson}}, \bibinfo
  {author} {\bibfnamefont {K.}~\bibnamefont {Cicak}}, \bibinfo {author}
  {\bibfnamefont {A.}~\bibnamefont {Metelmann}}, \bibinfo {author}
  {\bibfnamefont {S.}~\bibnamefont {Kotler}}, \bibinfo {author} {\bibfnamefont
  {R.}~\bibnamefont {Simmonds}}, \bibinfo {author} {\bibfnamefont
  {J.}~\bibnamefont {Teufel}},\ and\ \bibinfo {author} {\bibfnamefont
  {J.}~\bibnamefont {Aumentado}},\ }\bibfield  {title} {\bibinfo {title}
  {Microwave measurement beyond the quantum limit with a nonreciprocal
  amplifier},\ }\href {https://doi.org/10.1103/PhysRevApplied.13.044005}
  {\bibfield  {journal} {\bibinfo  {journal} {Phys. Rev. Appl.}\ }\textbf
  {\bibinfo {volume} {13}},\ \bibinfo {pages} {044005} (\bibinfo {year}
  {2020})}\BibitemShut {NoStop}%
\bibitem [{\citenamefont {Marques}\ and\ \citenamefont
  {Gross}(2004)}]{Marques2004}%
  \BibitemOpen
  \bibfield  {author} {\bibinfo {author} {\bibfnamefont {M.~A.}\ \bibnamefont
  {Marques}}\ and\ \bibinfo {author} {\bibfnamefont {E.~K.}\ \bibnamefont
  {Gross}},\ }\bibfield  {title} {\bibinfo {title} {Time-dependent density
  functional theory},\ }\href@noop {} {\bibfield  {journal} {\bibinfo
  {journal} {Annu. Rev. Phys. Chem.}\ }\textbf {\bibinfo {volume} {55}},\
  \bibinfo {pages} {427} (\bibinfo {year} {2004})}\BibitemShut {NoStop}%
\bibitem [{\citenamefont {Shirley}(1965)}]{Shirley1965}%
  \BibitemOpen
  \bibfield  {author} {\bibinfo {author} {\bibfnamefont {J.~H.}\ \bibnamefont
  {Shirley}},\ }\bibfield  {title} {\bibinfo {title} {Solution of the
  {S}chr\"odinger equation with a {H}amiltonian periodic in time},\ }\href
  {https://doi.org/10.1103/PhysRev.138.B979} {\bibfield  {journal} {\bibinfo
  {journal} {Phys. Rev.}\ }\textbf {\bibinfo {volume} {138}},\ \bibinfo {pages}
  {B979} (\bibinfo {year} {1965})}\BibitemShut {NoStop}%
\bibitem [{\citenamefont {Breuer}\ and\ \citenamefont
  {Petruccione}(2002)}]{Breuer2002}%
  \BibitemOpen
  \bibfield  {author} {\bibinfo {author} {\bibfnamefont {H.-P.}\ \bibnamefont
  {Breuer}}\ and\ \bibinfo {author} {\bibfnamefont {F.}~\bibnamefont
  {Petruccione}},\ }\href@noop {} {\emph {\bibinfo {title} {The theory of open
  quantum systems}}}\ (\bibinfo  {publisher} {Oxford University Press, USA},\
  \bibinfo {year} {2002})\BibitemShut {NoStop}%
\bibitem [{\citenamefont {Jiang}\ \emph {et~al.}(2024)\citenamefont {Jiang},
  \citenamefont {Gudem}, \citenamefont {Kowalewski},\ and\ \citenamefont
  {Dorfman}}]{Jiang2024}%
  \BibitemOpen
  \bibfield  {author} {\bibinfo {author} {\bibfnamefont {S.}~\bibnamefont
  {Jiang}}, \bibinfo {author} {\bibfnamefont {M.}~\bibnamefont {Gudem}},
  \bibinfo {author} {\bibfnamefont {M.}~\bibnamefont {Kowalewski}},\ and\
  \bibinfo {author} {\bibfnamefont {K.}~\bibnamefont {Dorfman}},\ }\bibfield
  {title} {\bibinfo {title} {Multidimensional high-harmonic echo spectroscopy:
  Resolving coherent electron dynamics in the euv regime},\ }\href
  {https://doi.org/10.1073/pnas.2304821121} {\bibfield  {journal} {\bibinfo
  {journal} {Proceedings of the National Academy of Sciences}\ }\textbf
  {\bibinfo {volume} {121}},\ \bibinfo {pages} {e2304821121} (\bibinfo {year}
  {2024})}\BibitemShut {NoStop}%
\bibitem [{\citenamefont {Ludlow}\ \emph {et~al.}(2015)\citenamefont {Ludlow},
  \citenamefont {Boyd}, \citenamefont {Ye}, \citenamefont {Peik},\ and\
  \citenamefont {Schmidt}}]{Ludlow2015}%
  \BibitemOpen
  \bibfield  {author} {\bibinfo {author} {\bibfnamefont {A.~D.}\ \bibnamefont
  {Ludlow}}, \bibinfo {author} {\bibfnamefont {M.~M.}\ \bibnamefont {Boyd}},
  \bibinfo {author} {\bibfnamefont {J.}~\bibnamefont {Ye}}, \bibinfo {author}
  {\bibfnamefont {E.}~\bibnamefont {Peik}},\ and\ \bibinfo {author}
  {\bibfnamefont {P.~O.}\ \bibnamefont {Schmidt}},\ }\bibfield  {title}
  {\bibinfo {title} {Optical atomic clocks},\ }\href
  {https://doi.org/10.1103/RevModPhys.87.637} {\bibfield  {journal} {\bibinfo
  {journal} {Rev. Mod. Phys.}\ }\textbf {\bibinfo {volume} {87}},\ \bibinfo
  {pages} {637} (\bibinfo {year} {2015})}\BibitemShut {NoStop}%
\bibitem [{\citenamefont {DeMille}\ \emph {et~al.}(2024)\citenamefont
  {DeMille}, \citenamefont {Hutzler}, \citenamefont {Rey},\ and\ \citenamefont
  {Zelevinsky}}]{DeMille2024}%
  \BibitemOpen
  \bibfield  {author} {\bibinfo {author} {\bibfnamefont {D.}~\bibnamefont
  {DeMille}}, \bibinfo {author} {\bibfnamefont {N.~R.}\ \bibnamefont
  {Hutzler}}, \bibinfo {author} {\bibfnamefont {A.~M.}\ \bibnamefont {Rey}},\
  and\ \bibinfo {author} {\bibfnamefont {T.}~\bibnamefont {Zelevinsky}},\
  }\bibfield  {title} {\bibinfo {title} {Quantum sensing and metrology for
  fundamental physics with molecules},\ }\href
  {https://doi.org/10.1038/s41567-024-02499-9} {\bibfield  {journal} {\bibinfo
  {journal} {Nature Physics}\ }\textbf {\bibinfo {volume} {20}},\ \bibinfo
  {pages} {741} (\bibinfo {year} {2024})}\BibitemShut {NoStop}%
\bibitem [{\citenamefont {Cerrillo}\ \emph {et~al.}(2021)\citenamefont
  {Cerrillo}, \citenamefont {Hays}, \citenamefont {Fatemi},\ and\ \citenamefont
  {Yeyati}}]{Cerrillo2021}%
  \BibitemOpen
  \bibfield  {author} {\bibinfo {author} {\bibfnamefont {J.}~\bibnamefont
  {Cerrillo}}, \bibinfo {author} {\bibfnamefont {M.}~\bibnamefont {Hays}},
  \bibinfo {author} {\bibfnamefont {V.}~\bibnamefont {Fatemi}},\ and\ \bibinfo
  {author} {\bibfnamefont {A.~L.}\ \bibnamefont {Yeyati}},\ }\bibfield  {title}
  {\bibinfo {title} {Spin coherent manipulation in josephson weak links},\
  }\href {https://doi.org/10.1103/PhysRevResearch.3.L022012} {\bibfield
  {journal} {\bibinfo  {journal} {Phys. Rev. Res.}\ }\textbf {\bibinfo {volume}
  {3}},\ \bibinfo {pages} {L022012} (\bibinfo {year} {2021})}\BibitemShut
  {NoStop}%
\bibitem [{\citenamefont {Strasberg}\ \emph {et~al.}(2022)\citenamefont
  {Strasberg}, \citenamefont {Modi},\ and\ \citenamefont
  {Skotiniotis}}]{Strasberg2022}%
  \BibitemOpen
  \bibfield  {author} {\bibinfo {author} {\bibfnamefont {P.}~\bibnamefont
  {Strasberg}}, \bibinfo {author} {\bibfnamefont {K.}~\bibnamefont {Modi}},\
  and\ \bibinfo {author} {\bibfnamefont {M.}~\bibnamefont {Skotiniotis}},\
  }\bibfield  {title} {\bibinfo {title} {How long does it take to implement a
  projective measurement?},\ }\href {https://doi.org/10.1088/1361-6404/ac5a7a}
  {\bibfield  {journal} {\bibinfo  {journal} {European Journal of Physics}\
  }\textbf {\bibinfo {volume} {43}},\ \bibinfo {pages} {035404} (\bibinfo
  {year} {2022})}\BibitemShut {NoStop}%
\end{thebibliography}%

\end{document}